# Optical and plasmonic properties of high electron density epitaxial and oxidative controlled titanium nitride thin films

Ikenna Chris-Okoro[1], Sheilah Cherono[1], Wisdom Akande[1], Swapnil Nalawade[2], Mengxin Liu[1], Catalin Martin[3], Valentin Craciun[1,4], R. Soyoung Kim[5], Johannes Mahl[5], Tanja Cuk[6], Junko Yano[5], Ethan Crumlin[5], J. David Schall[1], Shyam Aravamudhan[2], Maria Diana Mihai[7,8], Jiongzhi Zheng[9], Lei Zhang[9], Geoffroy Hautier[9], and Dhananjay Kumar[1*]

[1]Department of Mechanical Engineering, North Carolina A&T State University, Greensboro, NC 27411, USA

[2]Joint School of Nanoscience and Nanoengineering, North Carolina A &T State University, Greensboro, NC, USA 27401

[3]School of Theoretical & Applied Sciences, Ramapo College of New Jersey, Mahwah, NJ 07430, USA

[4]National Institute for Laser, Plasma, and Radiation Physics and Extreme Light Infrastructure for Nuclear Physics, RO 060042, Magurele, Romania

[5]Lawrence Berkeley National Laboratory, Berkeley, CA 94720

[6]Department of Chemistry, University of Colorado, Boulders, CO 80309, USA

[7]Horia Hulubei National Institute for Physics and Nuclear Engineering, Măgurele, IF, 077125, Romania

[8]Department of Physics, National University of Science and Technology Politehnica Bucharest, RO, 060042, Romania

[9]Department of Materials Science and Engineering, Dartmouth College, NH, USA 03755

*Corresponding Author Email: dkumar@ncat.edu

**Abstract**

The present work is focused on designing and synthesizing negative-permittivity, high melting point, mechanically hard, and chemically stable thin film materials beyond commonly employed plasmonic noble metals. The materials studied here are titanium nitride (TiN) and its isostructural rocksalt oxide derivative, titanium oxynitrides (TiNO) in thin film geometry. The advantages of oxide derivatives of TiN are the continuation of similar free electron density as in TiN and the acquiring of additional features such as oxygen dependent semiconducting bandgaps and optical reflectivity values, opening a new dimension to semiconducting plasmonics research. This work reports a pulsed laser-assisted synthesis, detailed structural characterization and study of plasmonic properties of three sets of TiN/TiNO thin films with high electron density. The first two sets of TiN films were grown at 600°C and 700°C under a high vacuum condition (≤2 ×10-6Torr). The third set of TiN film was grown in the presence of 5 mTorr of molecular oxygen at 700 °C. The purpose of making these three sets of TiN/TiNO films was to understand the role of film crystallinity and the role of the oxygen content of TiN films on their optical and plasmonic properties. The results have shown that TiN films deposited in a high vacuum are metallic, have large optical reflectance, and have high optical and electrical conductivity. The TiN films, grown in 5 mTorr $O_2$, were found to be partially oxidized and semiconducting with room temperature resistivity nearly three times larger than those of the TiN films grown under high vacuum conditions. The optical conductivity of these films was analyzed using a Kramers-Kronig transformation of reflectance and a Lorentz-Drude model; the optical conductivity determined by these two different methods agreed very well. To corroborate our experimental spectral observations, we have calculated the phonon dispersions and Raman active modes of TiNO using the virtual crystal approximation. A comparative analysis of the phonon dispersions between rutile $TiO_2$ and rocksalt TiNO has shown that the incorporation of nitrogen atoms does not significantly alter the phonon dispersions of rutile $TiO_2$. However, it results in the emergence of new phonon modes at approximately 7.128 THz (237.65 $cm^{-1}$) at the Gamma point, which corresponds to the experimentally observed Multi-Photon Phase-MPP (240 $cm^{-1}$-R). From the collateral study of experimental results and theoretical corroboration, a suitable multi-layer optical model was proposed for the TiN/TINO epitaxial thin films to extract the individual complex dielectric function from which many other optical parameters can be calculated.

## 1. Introduction

Harnessing the time independent interaction between light and matter in nonlinear optical processes is essential for various applications such as photocatalysis, photovoltaics, plasmonics and the identification of molecules [1-4]; and several materials and material synthesis strategies have been proposed for these applications. Nevertheless, it remains essential to enhance the inherent luminescent [5-7] and plasmonic properties [8-10] for better overall efficient energy utilization [11, 12]. Unlike conventional optics, plasmonics has enabled extraordinary concentration and enhancement of electromagnetic radiation beyond the diffraction limit of light. Though the coupling of light to electronic charge in certain metals has been investigated for decades, the advancement in Surface Enhanced Raman Spectroscopy (SERS) in the last 25 years has improved our knowledge of surface plasmons, which has the ability to confine light to specific wavelengths [13-15]. These plasmons are oscillations of the free electron gas system, often at optical frequencies but confined to the metal surface, rather than having collective oscillation of a gas[16]. By engineering the material structure, surface periodicity, oxidation state, etc., the resonance of the surface plasmons can be better controlled [16-18]. It has been shown that enhancing the performance of plasmonic materials has led to the development of nanometer-scale device sizes and THz operational bandwidths [19-23].

The oscillation of the free charge carriers, usually in noble metals such as Ag and Au provide the opportunity for tuning the plasmonic properties [23-27]. However, a major challenge of these noble metals is their mechanical softness and low melting point, especially when working in harsh operational conditions [21]; the next best alternative seems to be refractory metals. However, they lack plasmonic properties in the visible range and exhibit poor resonances in the near-infrared region [25]. Besides materials, structure, periodicity, and oxidation states, the shape and architecture of materials also have a profound effect on the properties of these materials [28-40]. For example, in nanowire form, the permittivity (ε) is different along the axes parallel or perpendicular to the propagation of light, where $\varepsilon_\parallel < 0, \epsilon_\perp > 0$. This relationship is reversed for thin-film materials, where $\varepsilon_\parallel > 0, \epsilon_\perp < 0$. When signs of the two components of permittivity are opposite, the isofrequency contours are unbounded and result in regular hyperbola for 2D thin film configuration as opposed to inverted hyperbolas in the case of 1D nanowires [10]. Thus, by balancing the frequency-dependent permittivities of the dielectric (positive) and metallic materials (negative) together with the geometrical parameters of the real part of $\varepsilon_{\parallel}$ and $\varepsilon^{\perp}$, one can obtain $\varepsilon_{\parallel}$ and $\varepsilon_{\perp}$ of opposite signs in the same materials.

The present work reports the plasmonic properties of titanium nitride (TiN) and titanium oxynitride (TiNO) [41, 42]. The general formula of the TiNO homolog series is expressed as $Ti^{3+}_{(1-\frac{x}{3})}Ti(vac)^{3+}_{\frac{x}{3}}N^{3-}_{(1-x)}O^{2-}_{x}$. The series exists in the whole range of $0 \leq x \leq 1$. In this formula, $Ti^{3+}$ (vac) is the number of $Ti^{3+}$ vacancies in each formula unit. This molecular formula takes into account the substitution of trivalent N anions by bivalent O anions in an ionic TiN lattice and the maintenance of charge neutrality in the lattice. The formula is also built on the assumption that the valences of Ti, N, and O are maintained at +3, -3, and -2, respectively. The terminal compounds (TiN with x=0 and TiO with x=1), as well as all the intermediate compounds with x between 0 and 1, all have rock salt structures, but they possess wide-ranging physicochemical properties.

TiN is well known to possess free electron gas density similar to that of Au or Ag and is the embodiment of refractory metals' characteristics[43, 44]. The availability of free electrons in TiN is brought about by the electronic configurations of ions involved and the bonding between them [45]. As seen in Figure 1, the TiN material system exhibits high-quality plasmonic resonances in the visible region and bears good refractory properties.

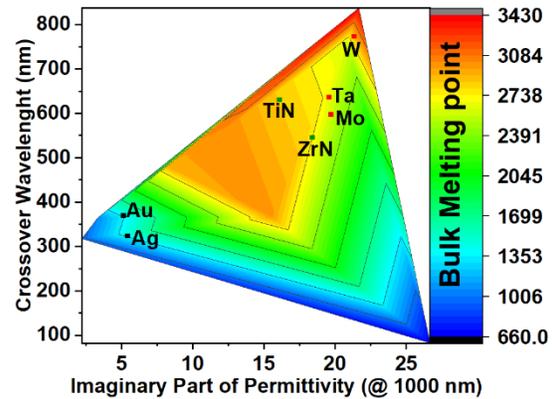

**Figure 1:** Comparison of selected transition metal nitrides (e.g., TiN, ZrN) with plasmonic noble metals (e.g., Au, Ag) and refractory (e.g., W, Ta, Mo) metals [28].

The other material advantages offered by TiN that enable new optoelectronic properties and device physics are: (i) high carrier concentration suitable for high plasma frequency ($\omega_p = (Ne^2/\epsilon_0 m_e)^{1/2}$) and tunable bandgaps to reduce interband transition losses, (ii) unusual combination of ionic, covalent, and metallic properties, (iii) ultrahigh hardness (close to that of diamond) and high melting point, (iv) brittleness and high thermal and electrical conductivity (higher than that of elemental transition metals). The aforementioned advantages of TiN are taken to the next level by transforming it (i.e. TiN) to oxynitrides (TiNO) with a precise control in oxygen composition that can, in turn, be used to tune the electronic band structure of transition metal nitrides (TMN), opening another new dimension to transition metals nitride-based plasmonics and metamaterials research. There is a sizable literature on the plasmonic properties of TMNs and their oxynitrides the interest in which continues to grow [46]. However, most of the research in this direction is targeted at producing plasmonic materials that have quality factors of localized surface plasmon resonances as close as possible to that of gold, silver, and copper. Some studies have focused on the operational temperature effects across various thicknesses [47], thickness effect [48-50], doping effect [51], effect of material geometries [52] on the plasmonic characteristics. In this study, the effect of varying oxidation state and crystallinity on the plasmonic performance of TiN and TiNO thin films is presented with theoretical calculations performed to understand how these fundamentals affect the overall performance of transitional metal nitrides and oxynitrides.

## 2. Results and Discussion

### 2.1. Structural Properties

The XRD diffraction patterns recorded from TiNO thin films grown on c-plane single-crystal sapphire substrates in a vacuum of 1.5×10$^{-6}$ Torr with no intentional addition of oxygen at 600°C and 700°C and in an oxygen pressure of 5 mTorr at 700 °C are shown in Figure 2a. All the films exhibit characteristic peaks of a rocksalt TiN crystal structure marked by first harmonic (111) peak at ~36.73° and second harmonic (222) peak at ~ 77.62° [44, 45, 53]. The sharpness of the XRD peaks is similar to the sharpness of a single crystal substrate peak (41.72°), and small full width at half-maximum (FWHM) values of the rocking curves (Figure 2b) indicate the epitaxial growth of TiN films, which was also confirmed by phi scans. As seen in Figure 2b, a high degree of crystallinity of these films is reflected from the Omega Rocking Curves (ORCs) for the (111) peaks. The FWHMs of the rocking curves of TiN films deposited in vacuum at 700°C and 600°C were found to be 0.154° and 0.364° respectively, while the FWHM for the TiN thin film deposited in 5 mTorr of O$_2$ at 700°C was considerably larger (0.716°). A double-decker shape of the rocking curve for the 5 mTorr of O$_2$ at 700°C sample could mean the presence of a mixed-phase material system, which is the case, as will be discussed later using XPS and Raman spectroscopy results. As a reference, the rocking curve FWHM of the sapphire (0001) peak was recorded to be 0.030°. While these films are textured and highly crystalline, there is a noticeable shift in the peak positions for all three films with respect to pure bulk TiN material [54]. The peak position shift is largest for the 600 °C, and least for the vacuum, 700°C sample. The shift in the peak positions is explained by the partial oxidation of TiN to TiNO, which also has the rocksalt crystal structure, but with a smaller cell parameter. The lattice constants of these films, calculated from the XRD-measured d-values, are 4.235 Å (vacuum, 700 °C), 4.189 Å (vacuum, 600 °C), 4.206 Å (5 mTorr oxygen, 700 °C). A decrease in the film lattice constant with a reduction in the deposition temperature and an increase in oxygen deposition pressure is thought to be associated with the smaller ionic radius of $O^{2-}$ (1.42 Å) than that of $N^{3-}$ (1.71 Å) [53, 54].

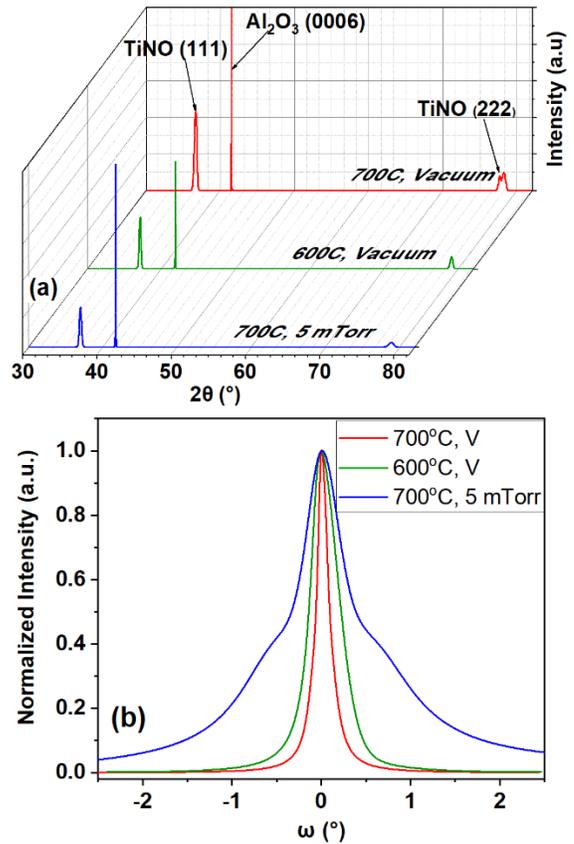

**Figure 2**. (a) X-ray diffraction patterns of TiNO deposited at different temperatures in a vacuum and 5 mTorr oxygen pressure, and (b) Omega Curves of the films whose XRD patterns are shown in (a).

Shown in Figure 3a is the phi-scan recorded to established the in-plane epitaxial relationship between the film and substrate with the rotation axis perpendicular to the film surface determined as $(111)_{TiNO}//(0001)_{Al_2O_3}$ and $<1\bar{1}0>_{TiNO} // <10\bar{1}0>_{Al_2O_3}$ [33, 44, 54, 55]. The observed reflections are TiNO(200) and $Al_2O_3$(102) plane.

For reference, projections of the atomic arrangement of (0001) $Al_2O_3$ and (111) TiN planes are provided in Figure 3b and Figure 3c, respectively. The $Al_2O_3$ structure coordinate file was obtained from Materials Project mp-1143 [56]. Atomistic renderings were produced using VESTA [57]. The reflection of TiNO film peaks spaced at 60° in the ϕ-scan is a characteristic of 6-fold symmetry (200 has 3-fold symmetry, the other peaks are coming from other orientations), while the reflection of $Al_2O_3$ film peaks spaced at 120° in the ϕ-scan is a characteristic of 3-fold symmetry. The ϕ-scan shows a ~ 30° rotation of the (111) film plane with respect to the (0001) plane of the substrate peaks. The 30° rotation is explained using 2-D oblique primitive unit cell structures of sapphire ($Al_2O_3$) and film (TiN) materials as illustrated in Figure 3d. For illustration purposes, the model sapphire surface is assumed to be oxygen terminated. This structure is representative of a Gibbsite-like $Al(OH)_3$ surface layer commonly found on sapphire exposed to ambient conditons which has been dehydrated due to exposure to elevated temperature during processing in vacuum before deposition [58, 59]. It should be noted that in the absence of environmental exporure, sapphire is typically Al terminated [59]. The film growth is assumed to begin with a titanium layer with titanium atoms located over hollowsites on the sapphire surface. The exact nature of the sapphire surface and TiN interface will be the subject of future theoretical studies. The 2D lattice parameters of sapphire and TiN are 4.8 Å and 3.0 Å, respectively. A 30° rotation of the TiN 2D cell with respect to the 2D cell of the sapphire substrate gives a 2D lattice parameter match of ~ 8.1%, which is less than the maximum accepted value of lattice mismatch of 10% that still allows epitaxial film growth. In this calculation, three 2D unit cells of TiN are matched with one 2D unit cell of sapphire (Figure 3d) via domain epitaxy[43, 44, 55]. This rotation is explained in terms of the TiNO layer formation and straining at the interface with the substrate [33, 54, 60, 61] due to alignment and rotation of the nitrogen or oxygen atoms of the TiNO film with respect to the oxygen atoms of the $Al_2O_3$ substrate leading to an energetically preferred orientation.

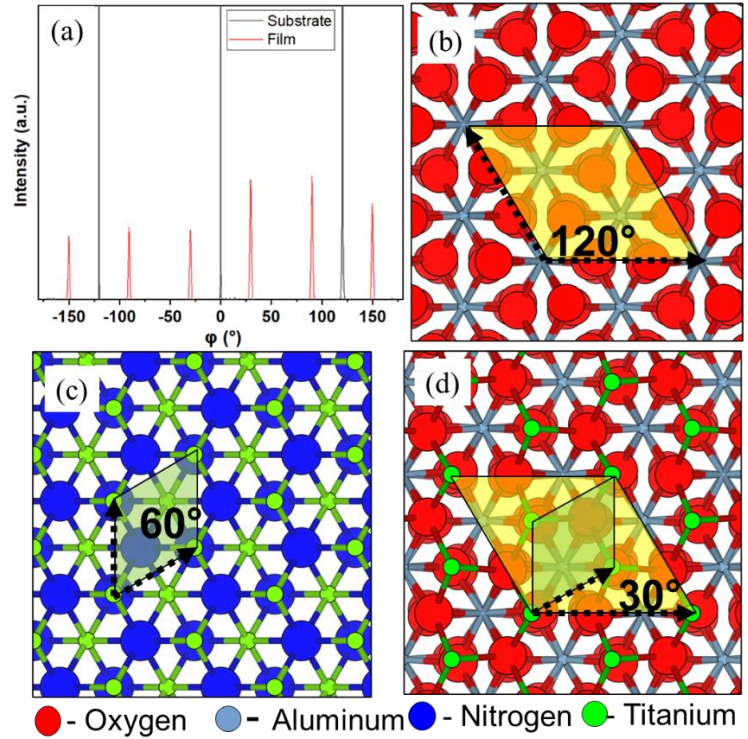

**Figure 3**: (a) φ-scan of a vacuum, 600°C TiNO-film-sapphire structure showing 30° rotation of the (111) film plane with respect to the (0001) plane of the substrate. (b) 2D primitive unit cell in the (0001) plane of sapphire, (c) 2D primitive unit cell in the (111) plane of TiN, (d)30° rotational matching of three 2D unit cells of TiN with one 2D unit cell of sapphire substrate.

The Ti 2p core level XPS spectrum is shown in Figure 4a. It is a common practice in the XPS analysis of the Ti 2p spectrum to assign unresolved and overlapping XPS features of differing shapes and intensities as satellite, shoulder, or hump [62-64]. For the deconvolution of the Ti 2p XPS spectra, a feature between the 456-459 eV is defined as a shake-up satellite [62]. A shake-up is caused by the transitions in the valence band that occur with a reduction of the kinetic energy from the emitted transitional metal band electron [63], others have argued against the in-situ formation of titanium dioxide and other oxynitride species [65, 66]. However, these approaches neglect the likeliness of the formation of various oxidation states of titanium that are common among transitional metal XPS spectra [67, 68]. As seen in Figure 4a, a doublet of Ti2p peaks is noticed in the XPS spectra, which is attributed to spin-orbit coupling beteeen Ti $2p_{3/2}$ and Ti $2p_{1/2}$. In our analysis, Ti 2p spectra were deconvolved to characterize Ti-N, Ti-N-O (as well as their plasmonic features), and Ti-O ($TiO_2$) species in all three samples, shown in Figure 4a. From the XPS spectra measured for the TiNO film deposited under various conditions, the peak position and FWHM of TiN $2p_{3/2}$ and $2p_{1/2}$, TiNO $2p_{3/2}$ and $2p_{1/2}$, and TiO$_2$ $2p_{3/2}$ and $2p_{1/2}$ were determined (as seen in Figure S1 and Table S2 and S3) which match well with those reported in the literature [39, 42, 62-73]. The varation in the FWHM of TiO$_2$ from the values reported in the literature is believed to be arising due to the deviation of TiO$_2$ stoichiometry (1.04 eV FHWM at 20 eV pass energy at Ti $2p_{3/2}$)[68]. The appropriateness of our fitting approach is illustrated by the fact that the interval between the $2p_{3/2}$ and $2p_{1/2}$-doublet peaks remained unchanged for TiN (5.9 eV) and TiO$_2$ (5.7 eV) for all the samples (Table S2). In contrast, this interval for TiNO varied from 5.72 to 5.82 as a function of film deposition conditions (Table S2). Also

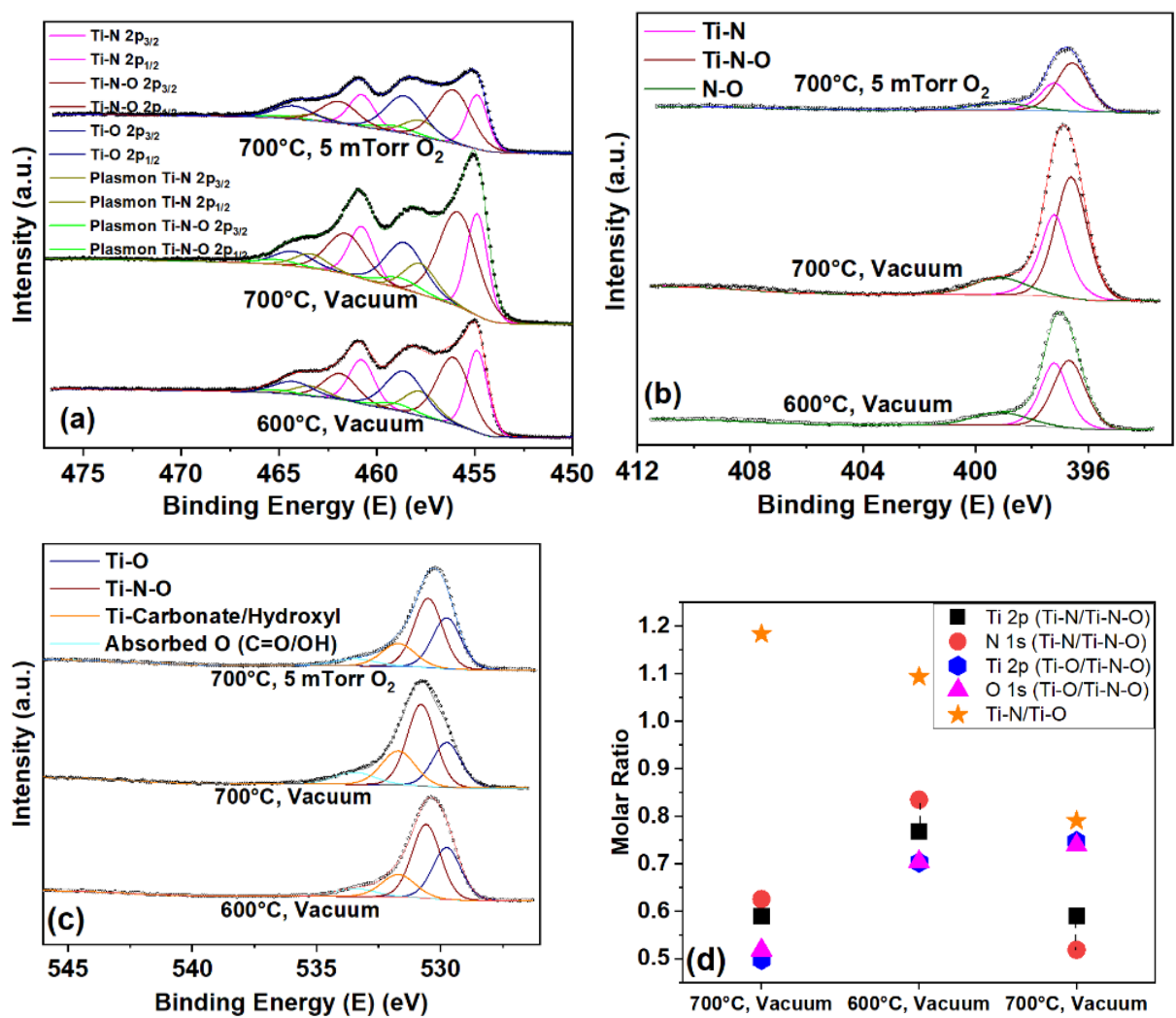

**Figure 4**. The XPS High-Resolution Spectra after 120s Ar Sputtering of (a) the Ti 2p core level, (b) the N 1s core level, and (c) the O 1s core level. (d) Molar Ratio comparision across the various spectra (Accuracy of Fit/Error bar).

In Figure 4b, the core-level N 1s spectrum has been deconvoluted into three peaks corresponding to titanium-nitrogen-oxygen bonding in TiNO, titanium-nitrogen bonding in TiN, and unattached nitrogen-oxygen referred to as chemisorbed nitrogen, respectively [39, 42, 62]. A mixture of surface effects (substitutional and interstitial nitrogen), which represents a variety of species such as N-O, N=O, and other nitrates accounts for the large FWHM as seen Table S4 and Figure S2. The molar fractions TiN and TiNO calculated from the deconvoluted N 1s peak agree with the relative molar fractions of TiN and TiNO calculated from the deconvoluted Ti2p peak (Figure 4d and Table S3). Similarly, the core-level O 1s spectrum has been deconvoluted into three peaks (Figure 4c) corresponding to titanium-oxygen-nitrogen bonding in TiNO, titanium-oxygen bonding in $TiO_2$ and (and possibly $Ti_2O_3$, $Ti_2O_5$, TiO), Ti-Carbonate/Hydroxyl bonding and adsorbed oxygen which represents a variety of species such as C-OH, C-O, C=O, N=O, and other atmospheric vapor oxides[39, 42, 62, 74] which accounts for the large FWHM as seen in Figure S3 and Table S5. Now, the molar fractions TiNO and $TiO_2$ calculated from the deconvoluted O1s peak agree with the molar fractions of TiNO and $TiO_2$ calculated from the deconvoluted Ti2p peak (Figure 4d and Table S3). The XPS results in reference to the elemental composition were confirmed using non-Rutherford Backscattering Spectrometry (Figure S4, Figure S5, Table S6 and Table S7).

Soft X-ray absorption spectroscopy (XAS) was carried out at the Ti $L_{3,2}$-edge, O K-edge, and N K-edge to further understand the chemical structure of the samples. The results obtained are presented in Figure 5. The spectra were acquired in total electron yield (TEY) mode, which is sensitive to the top <10 nm from the surface. The peaks in the Ti $L_{3,2}$-edge spectra (Figure 5a) correspond to excitation of Ti $2p_{3/2}$ and $2p_{1/2}$ electrons to empty Ti 3d states[75, 76]. The spectra shift to slightly higher energies in the order of 600 °C-vacuum → 700 °C-vacuum → 700 °C-5 mTorr $O_2$. The magnitude of the peaks, normalized to continuum absorption at 480 eV, also increases slightly in this order. This indicates increasing average oxidation state of Ti in the samples in this order [77]. This may seem contradictory with the XPS results that showed more oxygen in the 600 °C-vacuum sample than the 700 °C-vacuum sample. We note that XPS was acquired after $Ar^+$ sputtering while XAS was acquired as-is in surface-sensitive TEY mode; therefore, the 700 °C sample might be more oxidized on the top surface compared to the 600 °C sample. The O K-edge (Figure 5b) and N K-edge (Figure 5c) XAS show two peaks at low energy and broader peaks at higher energies. We attribute the pair of low-energy peaks to the excitation of O/N 1s electrons to O/N 2p orbitals hybridized with Ti 3d orbitals with $t_{2g}$ and $e_g$ symmetry, and the broad high-energy features to O/N 2p hybridized with Ti sp and other higher unoccupied states [78, 79]. The two vacuum samples have similar spectra at the O K-edge and N K-edge, implying similar average coordination environment around the anions. On the other hand, the 700 °C-5 mTorr $O_2$ sample had an altered spectrum, which is consistent with a significantly higher degree of oxidation. The N K-edge of this sample also showed a small peak on top of the $e_g$ peak at around 401 eV. This feature had been observed in other studies of oxidized TMNs and assigned to trapped $N_2$ molecules generated from oxidation of TiN.[76, 78, 80].

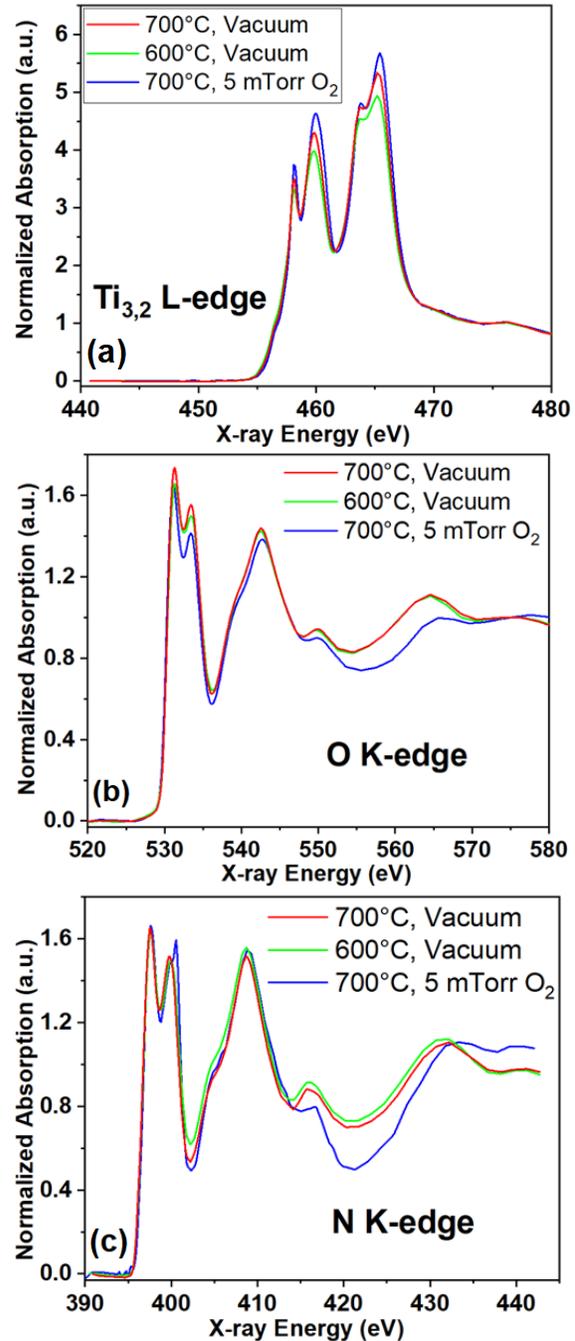

**Figure 5**. Soft X-ray absorption spectra of the (black) 600°C-Vacuum sample, (red) 700°C-Vacuum sample, and (blue) 700°C-5 mTorr $O_2$ sample at the (a) Ti $L_{3,2}$-, (b) O K-, and (c) N K-edge.

**2.2 Optical Properties**

The optical reflectance (R) characteristics of crystalline TiN films deposited at 600°C and 700°C under high vacuum conditions are shown in Figure 6a. A third set of TiN film was deposited in the presence of 5 mTorr of molecular oxygen at 700 °C. The purpose of testing the optical and plasmonic properties of these films was to understand the role of film crystallinity and the role of the oxygen content of TiN films that affects their conductivity. A large reflectance, at very low excitation energy (< 75 meV) and well-defined band edge between 22,000 cm$^{-1}$ - 25,000 cm$^{-1}$, observed for all samples, is consistent with the metallic behavior of these samples. Evidently, the 700°C- vacuum sample has the highest low frequency reflectance, R($\omega\rightarrow$0), followed by 600°C- vacuum sample and 700°C- 5 mTorr sample. The decrease in R($\omega\rightarrow$0) values indicates a decrease in dc-conductivity with oxygen content in the film. Accordingly, the position of the edge, associated with the plasma frequency of free carriers, shifts slightly to lower energies. The difference in the optical reflectance of these films can be interpreted using the XPS and XRD results.

According to the XRD and XPS results, the 700 °C - vacuum TiN sample is less oxidized and more crystalline than the 600 °C - vacuum TiN sample. Less oxygen content in the TiN film and therefore, possibly higher carrier concentration may explain the higher energy plasma edge in the 700 °C vacuum film. Then, the higher crystallinity may reduce the scattering rate, resulting in a higher electrical conductivity, hence higher low frequency reflectance, R($\omega\rightarrow$0)[81-91]. The intentional addition of oxygen gas during film growth at 700 °C is to oxidize the otherwise highly conducting TiN film to less conducting TiNO film. The XRD results have shown that this film has the largest FWHM among the three samples. These two effects are additively manifested in a decrease in the carrier concentration and an increase in the scattering rate at the same time. The trend is in good agreement with earlier presented results from the XPS data: the reduction of the TiN content results in lower free carrier concentration, while the reduction of crystallinity contributes to a higher scattering rate.

In order to clarify and quantify the optical behavior, further analysis of reflectance was carried out by performing separate measurements of the Kramers-Kronig transformation on the bare sapphire substrate and extracting complex optical functions of the films individually (Figure 6b). The analysis involved a model thin film of thickness $d_f$ on a thick substrate ($d_s$), therefore including three interfaces (air-film, film-substrate, and substrate-air) and four media, as sketched in the inset of Figure 6b. The first medium and the last medium are semi-infinite with a refractive index $n_0 = 1$. First, measured separately is the reflectance of the substrate and involved the using the Lorentz-Drude equation (Supporting Note S1) to fit the data and obtain a set of Drude ($\omega_p$, $1/\tau$) and Lorentz parameters ($\omega_j$, $S_j$, $1/\tau_j$) for the substrate. The fit was performed using the *dff* routine of the optical data analysis package *datan*, developed at the University of Florida [92], and the result is shown in Figure 6b. Using the obtained fitting parameters, we then calculated $\tilde{\varepsilon}_s$, $\tilde{n}_s = \sqrt{\tilde{\varepsilon}_s}$, and further determined the transfer matrix of the substrate $M_s$. Once the substrate is characterized, we use the *tff* routine for multi-layer structures of the same package, seeking to optimize a set of Lorentz-Drude parameters for the film ($\tilde{\varepsilon}_f$), while keeping fixed those of the substrate until the measured reflectance $\mathcal{R}_{meas}(\omega)$ is best reproduced. The results of the multi-layer structure fit are shown for each sample in Figure. 6a. It can be seen that the fit works well for the three samples.

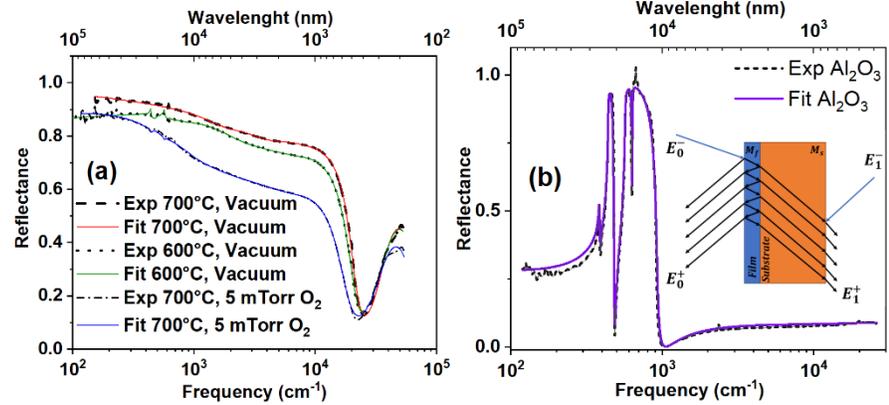

**Figure 6**: Reflectance spectra and the Lorentz-Drude fit of (a) the various film and (b) sapphire substrate (inset- a model of a thin film ($M_f$) on an Al$_2$O$_3$ substrate ($M_s$), at the air-film, film-substrate, and substrate-air interface and four media).

One of the main outcome of the multi-layer approach is that the individual complex dielectric function of film $\tilde{\varepsilon}_f$ is extracted, from which many other optical functions can be calculated (Figure S6). Figure 7a shows the real part of optical conductivity $\sigma_1(\omega) = \frac{\omega \varepsilon_2}{4\pi}$ of TiNO for all samples. The dc-conductivity $\sigma_{DC} = \sigma_1(0)$ confirms the qualitative trend with the film oxidation observed from direct measurements of reflectance, and at the same time, its values are more than double those obtained by neglecting the substrate effect, which agrees well with the direct dc-transport measurements

The dominant zero-frequency (Drude) peak and the strong absorption at high frequency between 33,000 cm$^{-1}$ to 37,000 cm$^{-1}$ (4.09 eV - 4.58eV) is observed as well as 2 additional spectral weights, below the absorption band of TiN [91] at the mid-infrared absorption 400 cm$^{-1}$ (0.05 eV) and 4000 cm$^{-1}$ (0.5 eV). To further confirm the metallic character of the TiNO films, shown in Figure 7(b) the loss function, $Loss = -Im(\frac{1}{\tilde{\varepsilon}})$, which is expected to have a peak at the plasma frequency, with its width related to the scattering rate of free carriers. Figure 7b clearly shows a peak is observed between the 19000 cm$^{-1}$ and 22,000 cm$^{-1}$ (2.36 eV- 2.73 eV) which the position of the band edge shifts to low excitation energy (related to increase in the band gap) and broadens as a function of plasma frequency of free carriers present in the various samples. As observed from the reflectance band edge is an indication of the reduction in concentration and enhanced scattering rate because of decrease in crystallinity. For good plasmonic material a large negative dielectric constant $\varepsilon_1$ becomes a necesity as is seen in Figure 7c. and the inset image for most of the spectra the most crystalline material (700 °C, Vacuum) has the largest negative values and, the largest frequency (energy) of zero-crossing. Important to note is that all samples show practicable plasmonic applications in the near to mid-infrared range

Aside from low losses, a large negative dielectric constant $\varepsilon_1$ is required for plasmonic devices, which has indeed been accomplished, as seen in Figure 7a for all films. The sample obtained in a vacuum has the largest negative values and the largest frequency (energy) of zero-crossing, which is the upper limit for use as plasmonic materials. Interestingly, even the oxidized TiN compounds (TiNO) show feasibility for plasmonic applications, at least in the near to mid-infrared range. The figure of merit (FOM), taken as the modulus of real ($\varepsilon_1$) and imaginary part ($\varepsilon_2$), for all samples are plotted in Figure 7d as a function of frequency and wavelength. For comparison, we have indicated the peak FOM for Au and Ag in the same graph. Au and Ag have nearly an order of higher magnitude FOM with respect to TiN/TiNO samples. However, the tunability of metallicity of TiN/TiNO by means of controlled oxidation and isomorphous phase transformation may offer advantages in terms of swifter conversion of light into heat/electrical energies and may find applications in photothermal and photocatalytic devices (Figure S6). As stoichiometric rocksalt TiN is highly symmetrical, its first-order Raman scattering is forbidden, and therefore, no active Raman mode should be observed [93]. However, the substitution of N by O breaks the symmetry, and several Raman peaks are visible as in Figure S7 which are similar to band of $TiO_2$ that are marked A and R for anatase and rutile $TiO_2$. The peak at 316 $cm^{-1}$ is attributed to the Raman Scattering at the Ti-N-O as there is no literature-reported value of $TiO_2$ around this range, as can be seen in Figure 8 [72].

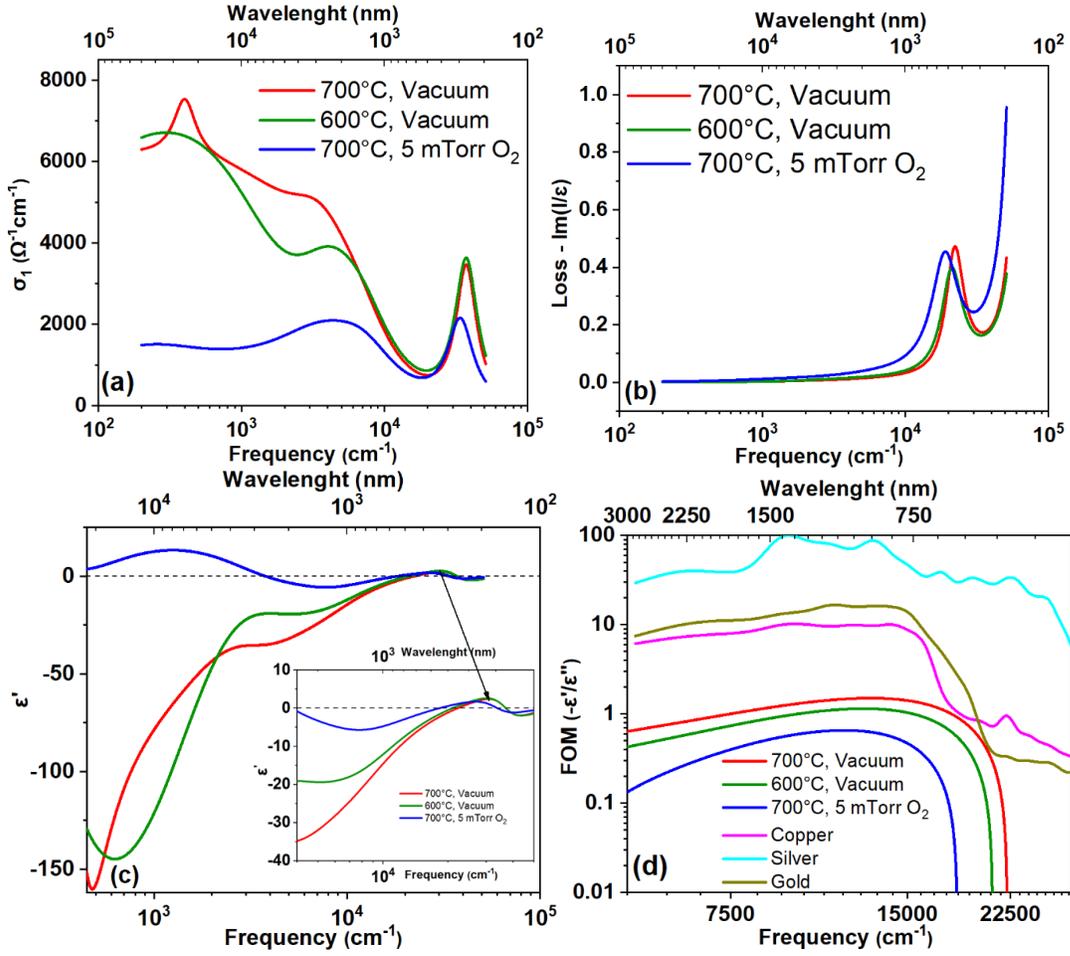

**Figure 7**(a) Optical Conductivity spectra from the Lorentz-Drude fit of the various film (b) loss function $Loss = -Im(\frac{1}{\varepsilon})$ (c) The real part of dielectric function $\varepsilon_1(\omega)$. (d) Figure of merit (FOM) $-\varepsilon'/\varepsilon''$ of the various films (compared with gold, copper, silver)[94].

This result compares very well with the XPS Ti 2p deconvolution discussed earlier in this study, where the formation of TiN and TiNO is much higher in concentration than $TiO_2$.

The Ti-N-O bond phonon vibrational modes can either be transversal (T), acoustic (A), Longitudinal (L), Optical (O) [72, 95], and more specifically, Longitudinal Acoustic (LA) at ~ 320 $cm^{-1}$,

Transverse Acoustic (TA) at ~ 235 cm$^{-1}$, and Longitudinal Optical (TO) at ~ 570 cm$^{-1}$ [96-101]. As Raman spectra mirror the density of vibrational states (DVS) [95], it is observed that the Raman Spectra is dominated by an asymmetric band centered at around 320 cm$^{-1}$ and in the lower frequency range by the presence of a phonon band centered at 230 cm$^{-1}$, which signify the density of vibrational states of TiNO films. The higher frequency range of the asymmetrical band is attributed to the superposed contributions of the disorder of acoustic phonons and 2$^{nd}$ order combination of optical and acoustic processes, while the lower frequency range of the asymmetrical band is attributed to the disorder of single phonon and other 2$^{nd}$ order processes which are not well defined [102-104].

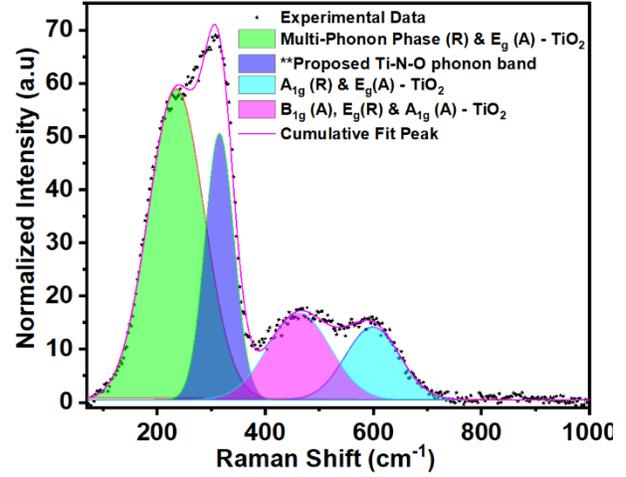

**Figure 8**: (Raman Spectra from the TiNO deposited at 700°C and 600°C at partial vacuum and 5 mTorr with 532 nm wavelength laser excitation. (b) peak fitting for TiNO with respect to the various phonon bands (700°C, 5mTorr).

To corroborate our experimental spectral observations, we have calculated the phonon dispersions and Raman active modes for anatase (A) and rutile (R) TiO$_2$, as presented in Figure 9a and Figure 9b. The irreducible representations of rutile (R) TiO$_2$ are B$_{1u}$(85.50 cm$^{-1}$), A$_{2u}$(100.81 cm$^{-1}$), B$_{1g}$(143.75 cm$^{-1}$), E$_u$(357.95 cm$^{-1}$), B$_{1u}$(370.31 cm$^{-1}$), A$_{2g}$(408.27 cm$^{-1}$), E$_g$(441.35 cm$^{-1}$), E$_u$(482.46 cm$^{-1}$), A$_{1g}$(582.13 cm$^{-1}$), and B$_{2g}$ (785.37 cm$^{-1}$). For anatase (A) TiO$_2$, the irreducible representations are E$_g$(116.03 cm$^{-1}$), E$_g$(171.01 cm$^{-1}$), E$_u$(211.56 cm$^{-1}$), A$_{2u}$(306.14 cm$^{-1}$), B$_{1g}$(364.29 cm$^{-1}$), E$_u$(398.50 cm$^{-1}$), B$_{1g}$(477.95 cm$^{-1}$), A$_{1g}$(496.26 cm$^{-1}$), B$_{2u}$(509.05 cm$^{-1}$), and E$_g$(602.68 cm$^{-1}$). Indeed, it is not possible to assign the Multi-Photon Phase-MPP (240 cm$^{-1}$-R) to any of the irreducible representations observed in either anatase (A) or rutile (R) TiO$_2$. We thus proceed to calculate the phonon dispersions of TiNO using the virtual crystal approximation. A comparative analysis of the phonon dispersions between rutile TiO$_2$ and TiNO is depicted in Figure 9c. Clearly, the incorporation of nitrogen atoms does not significantly alter the phonon dispersions of rutile TiO$_2$. However, it results in the emergence of new phonon modes at approximately 7.128 THz (237.65 cm$^{-}$ [1]) at the Gamma point, which corresponds to the experimentally observed Multi-Photon Phase-MPP (240 cm$^{-1}$-R).

## 3. Experimental method

Titanium nitride and titanium oxynitride thin film were deposited on a 10 mm×10 mm×0.5 mm c-plane sapphire (0001) substrate using a pulsed laser deposition (PLD) technique. A high-purity (99.99%) TiN target (one-inch diameter, quarter-inch thickness) was used to deposit the TiN/TiNO films by controlling the substrate temperature, oxygen pressure, laser energy density, and laser pulse repetition rate. A KrF laser (Lambda Physik, wavelength 248 nm, pulse duration 30 ns) was used. The substrates were cleaned using the method described in our earlier publications[32-35]. The PLD chamber was pumped down to a base pressure of <3 × 10$^{-6}$ Torr, and subsequently, the substrate heater was heated to the desired temperatures. A fixed number of laser pulses of 20,000 was used at a laser frequency of 10 Hz. The thin film deposition experiments were carried out in a 5 mTorr $O_2$ pressure and under a vacuum of 1.5×10$^{-6}$ Torr; the deposition in ~10$^{-6}$ Torr vacuum is referred to as deposition in residual oxygen ambient, i.e., no oxygen added intentionally. The substrate temperatures of 600°C to 700°C were used, keeping all other deposition parameter the same in all experiments.

The surface morphology of TiN and TiNO films was investigated using atomic force microscopy (Asylum Jupiter XR). To measure the elemental distribution of Ti, N, and O a, Hitachi SU8000 scanning electron microscopy (SEM) and energy dispersive x-ray spectroscopy (EDS) were used. The electrical resistivity was determined using an Ossila T2001A standard four-probe measurement. The unit lattice models were simulated using the Visualization for Electronic and Structural Analysis (VESTA). The film- orientation, thickness, and crystallinity were analyzed using the x-ray diffractometer (Rigaku Smartlab XRD) with a high-flux CuK$_\alpha$ X-ray source ($\lambda$ = 0.154 nm). XPS measurements were carried out using a Thermo Escalab Xi+ working with Al Ka monochromatic radiation. Raman Spectroscopy measurements were carried out using a WiTec alpha300R Confocal Raman Microscope at the laser excitation wavelength of 532 nm (green visible light).

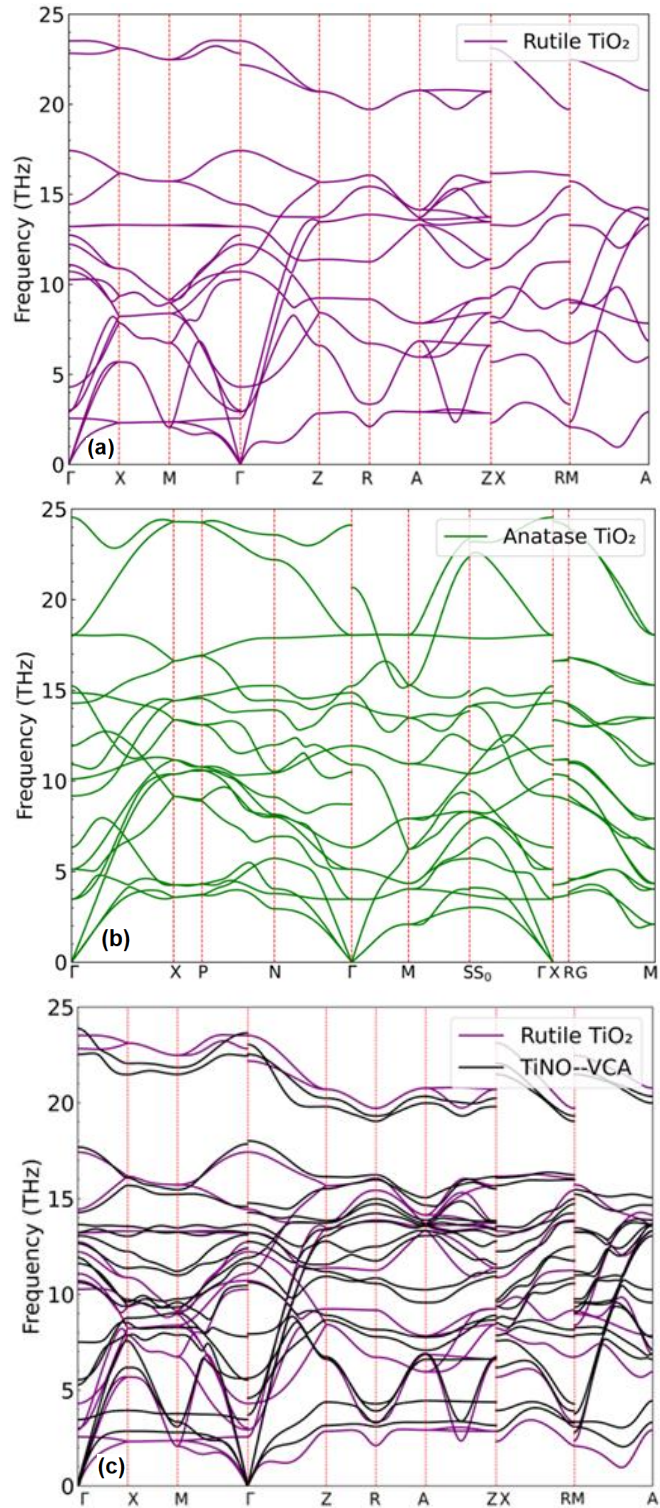

**Figure 9**: (a) Calculated Phonon dispersion of rutile TiO$_2$. (b) Calculated Phonon dispersions of anatase TiO$_2$. (c) Comparison of phonon dispersions between rutile TiO$_2$ and virtual crystal TiNO (frequency unit in THz).

High-resolution x-ray photoelectron spectroscopy (XPS) were recorded for Ti 2p, N 1s, and O 1s core levels to accurately quantify the oxidized, partially oxidized, and unoxidized phases of TiN by taking care of the common errors frequently encountered right from the data collection to subsequent analysis. A precise quantification of these phases is important in understanding the resulting properties of TiNO compounds formed at higher deposition temperatures and in the presence of oxygen ambient. In this respect, all XPS spectra were recorded under low noise with a reasonable peak-to-background ratio. The caution exercised in the subsequent data analysis involved selecting proper model functions, reducing the number of free parameters by means of existing information on well-known doublet splitting and intensity ratios, running the fit procedure with Shirley background subtraction with 50 eV pass energy and 30 eV for survey and high resolution scan respectively to obtain a good signal-noise ratio, and taking only < 7 units residual standard deviation fittings[69]. Also, in XPS analysis and fitting, binding energy, FWHM, and spectra shape are kept identical for the same chemistry. For example, the binding energies, FWHM, and spectra shape obtained using a set of fitting parameters should be independent of sample preparation conditions for one kind of chemistry, namely in the present study TiN and $TiO_2$ chemistry. However, a variation in these parameters (binding energies, FWHM, and spectra shape) for TiNO using the same fitting precautions used for TiN and $TiO_2$ with well-established chemistries should be taken to confirm changing chemistry and changing O/N ratio in the various films. XAS measurements were carried out at Beamline 7.3.1 at the Advanced Light Source (ALS), Lawrence Berkeley National Laboratory (LBNL).

The elemental composition of the TiN and TiNO films was also determined using non-Rutherford backscattering spectrometry (NRBS) with $^4He^{++}$ ions at 3.043 MeV and 3.7 MeV. Light elements like oxygen (O) and nitrogen (N) exhibit a higher cross-section at these specific energies, enabling their differentiation from the substrate signal. The NRBS measurements were performed under high vacuum ($10^{-6}$ mbar), using a collimated $^4He^{++}$ beam extracted from the duoplasmatron ion source of the 3 MV Tandetron accelerator of Horia Hulubei National Institute for R&D in Physics and Nuclear Engineering (IFIN-HH)[105-107]. The alpha particles were detected with a passivated, ion-implanted silicon detector positioned at an angle of 165° relative to the incident beam direction. With a detector diameter of 8 mm and a sample-to-detector distance of 175 mm, the configuration yielded a solid angle of 1.641 msr. The measured energy resolution of the detector was 17 keV. During NRBS, the samples were tilted at 7° relative to the beam direction. The total ion dose was 20 μC per spectrum. The analysis of recorded NRBS spectra was performed using the SIMNRA simulation code [106]. Ab-initio calculations based on density functional theory (DFT) using the Vienna Ab-initio Simulation Package (VASP) were carried out as is discussed in Supporting Note 3. To ensure the authenticity and reproducibility of the data presented in this study, samples were deposited twice under identical conditions, and then their properties (thickness, crystallinity, four-probe resistivity, Raman Spectra, NRBS, and XPS compositions) were measured at least twice.

4. Conclusions

The present study has carried out work on the thin film fabrication and precision characterization of titanium nitride and its isostructural oxidative derivative based negative-permittivity, high melting point, mechanically hard, and chemically stable materials beyond commonly employed plasmonic metals (e.g., Au, Ag). The advantages of TiN to possess free electron gas density similar to Au and Ag are taken to the next level by transforming TiN to oxynitrides with precise control in oxygen composition that can, in turn, be used to tune the electronic band structure of TMNs. An accurate determination of the variations of the molar fractions TiN and TiNO in the samples prepared in different oxidation conditions has been accomplished by observing a near perfect match in their fractions calculated from the N1s and the Ti2p peaks in the x-ray photoelectron spectroscopy (XPS). This accuracy is further confirmed by a near-perfect match in the molar fractions of TiNO and $TiO_2$ calculated from the XPS O1s and the Ti2p peaks. The optical conductivity of these films was analyzed using a Kramers-Kronig transformation of reflectance and a Lorentz-Drude model; the optical conductivity determined by two different methods agrees very well.

**AUTHOR INFORMATION**

**Corresponding Author**

**Dhananjay Kumar**- Department of Mechanical Engineering, North Carolina A&T State University, Greensboro, NC, USA; https://orcid.org/0000-0001-5131-5131; Email: dkumar@ncat.edu

**Authors**

**Ikenna Chris-Okoro**- Department of Mechanical Engineering, North Carolina A&T State University, Greensboro, NC, USA 27411

**Sheilah Cherono**- Department of Mechanical Engineering, North Carolina A&T State University, Greensboro, NC, USA 27411

**Wisdom Akande**- Department of Mechanical Engineering, North Carolina A&T State University, Greensboro, NC, USA 27411

**Swapnil Nalawade**- Joint School of Nanoscience and Nanoengineering, North Carolina A &T State University, Greensboro, NC, USA 27401

**Mengxin Liu**- Department of Mechanical Engineering, North Carolina A&T State University, Greensboro, NC, USA 27411

**Catalin Martin** - School of Theoretical & Applied Sciences, Ramapo College of New Jersey, Mahwah, NJ, USA 07430

**Valentin Craciun**- Department of Mechanical Engineering, North Carolina A&T State University, Greensboro, NC, USA 27411; National Institute for Laser, Plasma and Radiation Physics, Bucharest-Magurele, Romania 077125

**Soyoung Kim**- Lawrence Berkeley National Laboratory, Berkeley, CA, USA 94720

**Johannes Mahl**- Lawrence Berkeley National Laboratory, Berkeley, CA, USA 94720

**Tanja Cuk**- Department of Chemistry, University of Colorado, Boulders, CO, USA 80309

**Junko Yano**- Lawrence Berkeley National Laboratory, Berkeley, CA, USA 94720

**Ethan Crumlin**- Lawrence Berkeley National Laboratory, Berkeley, CA, USA 94720

**J. David Schall**- Department of Mechanical Engineering, North Carolina A&T State University, Greensboro, NC, USA 27411

**Shyam Aravamudhan**- Joint School of Nanoscience and Nanoengineering, North Carolina A &T State University, Greensboro, NC, USA 27401

**Maria Diana Mihai**-Horia Hulubei National Institute for Physics and Nuclear Engineering, Măgurele, IF, 077125, Romania; [7]Department of Physics, National University of Science and Technology Politehnica Bucharest, RO, 060042, Romania

**Jiongzhi Zheng**- Department of Materials Science and Engineering, Dartmouth College NH, USA 03755

**Lei Zhang**- Department of Materials Science and Engineering, Dartmouth College NH, USA 03755

**Geoffroy Hautier**-Department of Materials Science and Engineering, Dartmouth College NH, USA 03755

## Author Contributions

The manuscript was written through the contributions of all authors. All authors have given approval to the final version of the manuscript.

**Acknowledgments**

The research was supported by the Center for Electrochemical Dynamics and Reactions on Surfaces (CEDARS), an Energy Frontier Research Center, funded by the U.S. Department of Energy (DOE), Office of Science, Basic Energy Sciences (BES) via grant # DE-SC0023415.

Part of the work was performed using the resources at the Joint School of Nanoscience and Nanotechnology, a member of the National Nanotechnology Coordinated Infrastructure (NNCI), which is supported by the National Science Foundation (Grant ECCS-2025462), and resources built using NSF-PREM CREAM project at NCAT. VC work was funded by grants of the Romanian Ministry of Scientific Research, Innovation, and Digitalization project PN-III-P4-PCE-2021-1158 and Core Program LAPLAS VII 30N/2023. The work at the ALS of the LBNL was supported by the Director, Office of Science, Office of Basic Energy Sciences, of the US Department of Energy under Contract No. DE-AC02-05CH11231.

**Data availability statement**

The raw/processed data required to reproduce these findings cannot be shared at this time due to technical or time limitations.

**Declaration of Competing Interest**

The authors declare that they have no known competing financial interests or personal relationships that could have appeared to influence the work reported in this paper.

# References

[1] R. Gutzler, M. Garg, C. R. Ast, K. Kuhnke, and K. Kern, "Light–matter interaction at atomic scales," *Nature Reviews Physics,* vol. 3, no. 6, pp. 441-453, 2021.
[2] D. Maiti, J. Cairns, J. N. Kuhn, and V. R. Bhethanabotla, "Interface engineering of metal oxynitride lateral heterojunctions for photocatalytic and optoelectronic applications," *The Journal of Physical Chemistry C,* vol. 122, no. 39, pp. 22504-22511, 2018.
[3] N. Rivera and I. Kaminer, "Light–matter interactions with photonic quasiparticles," *Nature Reviews Physics,* vol. 2, no. 10, pp. 538-561, 2020.
[4] X. Yang *et al.*, "Nitrogen-doped ZnO nanowire arrays for photoelectrochemical water splitting," *Nano letters,* vol. 9, no. 6, pp. 2331-2336, 2009.
[5] G. K. Mor, K. Shankar, M. Paulose, O. K. Varghese, and C. A. Grimes, "Use of highly-ordered TiO2 nanotube arrays in dye-sensitized solar cells," *Nano letters,* vol. 6, no. 2, pp. 215-218, 2006.
[6] J. Nowotny, T. Bak, M. Nowotny, and L. Sheppard, "Titanium dioxide for solar-hydrogen I. Functional properties," *International journal of hydrogen energy,* vol. 32, no. 14, pp. 2609-2629, 2007.
[7] B. D. Alexander, P. J. Kulesza, I. Rutkowska, R. Solarska, and J. Augustynski, "Metal oxide photoanodes for solar hydrogen production," *Journal of Materials Chemistry,* vol. 18, no. 20, pp. 2298-2303, 2008.
[8] J. Lee, S. Mubeen, X. Ji, G. D. Stucky, and M. Moskovits, "Plasmonic photoanodes for solar water splitting with visible light," *Nano letters,* vol. 12, no. 9, pp. 5014-5019, 2012.
[9] L. G. Devi and R. Kavitha, "A review on plasmonic metal☐ TiO2 composite for generation, trapping, storing and dynamic vectorial transfer of photogenerated electrons across the Schottky junction in a photocatalytic system," *Applied Surface Science,* vol. 360, pp. 601-622, 2016.
[10] P. Wang *et al.*, "Molecular plasmonics with metamaterials," *Chemical Reviews,* vol. 122, no. 19, pp. 15031-15081, 2022.
[11] Y. Liu, Y. Li, W. Li, S. Han, and C. Liu, "Photoelectrochemical properties and photocatalytic activity of nitrogen-doped nanoporous WO3 photoelectrodes under visible light," *Applied Surface Science,* vol. 258, no. 12, pp. 5038-5045, 2012.
[12] T. S. Natarajan, V. Mozhiarasi, and R. J. Tayade, "Nitrogen doped titanium dioxide (N-TiO2): synopsis of synthesis methodologies, doping mechanisms, property evaluation and visible light photocatalytic applications," *Photochem,* vol. 1, no. 3, pp. 371-410, 2021.
[13] G. Barbillon, *Nanoplasmonics - Fundamentals and Applications*, 3 ed. 2017.
[14] M. I. S. Tigran V. Shahbazyan, Ed. *Plasmonics: Theory and Applications*, 1 ed. (Challenges and Advances in Computational Chemistry and Physics). Springer Dordrecht, 2013, pp. 1-577.
[15] G. Barbillon, "Plasmonics and its Applications," *Materials (Basel),* vol. 12, no. 9, May 8 2019.
[16] J.-F. Bryche *et al.*, "Density effect of gold nanodisks on the SERS intensity for a highly sensitive detection of chemical molecules," *Journal of Materials Science,* vol. 50, no. 20, pp. 6601-6607, 2015.
[17] Y. Lee *et al.*, "Particle-on-Film Gap Plasmons on Antireflective ZnO Nanocone Arrays for Molecular-Level Surface-Enhanced Raman Scattering Sensors," *ACS Appl Mater Interfaces,* vol. 7, no. 48, pp. 26421-9, Dec 9 2015.
[18] J.-F. Masson, K. F. Gibson, and A. Provencher-Girard, "Surface-Enhanced Raman Spectroscopy Amplification with Film over Etched Nanospheres," *The Journal of Physical Chemistry C,* vol. 114, no. 51, pp. 22406-22412, 2010.
[19] S. Lal, S. Link, and N. J. Halas, "Nano-optics from sensing to waveguiding," *Nature Photonics,* vol. 1, no. 11, pp. 641-648, 2007.
[20] A. Boltasseva and V. M. Shalaev, "All that glitters need not be gold," *Science,* vol. 347, no. 6228, pp. 1308-10, Mar 20 2015.

[21]  U. Guler, A. Boltasseva, and V. M. Shalaev, "Refractory plasmonics," *Science,* vol. 344, no. 6181, pp. 263-4, Apr 18 2014.

[22]  A. M. Shaltout, J. Kim, A. Boltasseva, V. M. Shalaev, and A. V. Kildishev, "Ultrathin and multicolour optical cavities with embedded metasurfaces," *Nat Commun,* vol. 9, no. 1, p. 2673, Jul 10 2018.

[23]  M. Usman Javed *et al.*, "Tailoring the plasmonic properties of complex transition metal nitrides: A theoretical and experimental approach," *Applied Surface Science,* vol. 641, 2023.

[24]  J. Li, S. K. Cushing, P. Zheng, F. Meng, D. Chu, and N. Wu, "Plasmon-induced photonic and energy-transfer enhancement of solar water splitting by a hematite nanorod array," *Nat Commun,* vol. 4, p. 2651, 2013.

[25]  J. A. Scholl, A. L. Koh, and J. A. Dionne, "Quantum plasmon resonances of individual metallic nanoparticles," *Nature,* vol. 483, no. 7390, pp. 421-7, Mar 21 2012.

[26]  Y. Gutiérrez, A. S. Brown, F. Moreno, and M. Losurdo, "Plasmonics beyond noble metals: Exploiting phase and compositional changes for manipulating plasmonic performance," *Journal of Applied Physics,* vol. 128, no. 8, 2020.

[27]  C. W. Moon, M. J. Choi, J. K. Hyun, and H. W. Jang, "Enhancing photoelectrochemical water splitting with plasmonic Au nanoparticles," *Nanoscale Adv,* vol. 3, no. 21, pp. 5981-6006, Oct 27 2021.

[28]  U. Guler *et al.*, "Local heating with lithographically fabricated plasmonic titanium nitride nanoparticles," *Nano Lett,* vol. 13, no. 12, pp. 6078-83, 2013.

[29]  L. Mascaretti, C. Mancarella, M. Afshar, S. Kment, A. L. Bassi, and A. Naldoni, "Plasmonic titanium nitride nanomaterials prepared by physical vapor deposition methods," *Nanotechnology,* vol. 34, no. 50, Oct 11 2023.

[30]  P. Subramanyam, B. Meena, V. Biju, H. Misawa, and S. Challapalli, "Emerging materials for plasmon-assisted photoelectrochemical water splitting," *Journal of Photochemistry and Photobiology C: Photochemistry Reviews,* vol. 51, 2022.

[31]  F. X. Xiao and B. Liu, "Plasmon-Dictated Photo-Electrochemical Water Splitting for Solar-to-Chemical Energy Conversion: Current Status and Future Perspectives," *Advanced Materials Interfaces,* vol. 5, no. 6, 2018.

[32]  K. Sarkar, S. Shaji, S. Sarin, J. E. Shield, C. Binek, and D. Kumar, "Large refrigerant capacity in superparamagnetic iron nanoparticles embedded in a thin film matrix," *Journal of Applied Physics,* vol. 132, no. 19, 2022.

[33]  M. Roy *et al.*, "Modulation of Structural, Electronic, and Optical Properties of Titanium Nitride Thin Films by Regulated In Situ Oxidation," *ACS Appl Mater Interfaces,* vol. 15, no. 3, pp. 4733-4742, Jan 25 2023.

[34]  N. R. Mucha *et al.*, "High-Performance Titanium Oxynitride Thin Films for Electrocatalytic Water Oxidation," *ACS Applied Energy Materials,* vol. 3, no. 9, pp. 8366-8374, 2020.

[35]  N. R. Mucha *et al.*, "Electrical and optical properties of titanium oxynitride thin films," *Journal of Materials Science,* vol. 55, no. 12, pp. 5123-5134, 2020.

[36]  I. R. Howell, B. Giroire, A. Garcia, S. Li, C. Aymonier, and J. J. Watkins, "Fabrication of plasmonic TiN nanostructures by nitridation of nanoimprinted TiO2 nanoparticles," *Journal of Materials Chemistry C,* vol. 6, no. 6, pp. 1399-1406, 2018.

[37]  K. Kamiya, T. Yoko, and M. Bessho, "Nitridation of TiO2 fibres prepared by the sol-gel method," *Journal of Materials Science,* vol. 22, no. 3, pp. 937-941, 1987.

[38]  P. Romero-Gómez, V. Rico, J. P. Espinós, A. R. González-Elipe, R. G. Palgrave, and R. G. Egdell, "Nitridation of nanocrystalline TiO2 thin films by treatment with ammonia," *Thin Solid Films,* vol. 519, no. 11, pp. 3587-3595, 2011.

[39]  M. Pisarek, M. Krawczyk, M. Holdynski, and W. Lisowski, "Plasma Nitriding of TiO(2) Nanotubes: N-Doping in Situ Investigations Using XPS," *ACS Omega,* vol. 5, no. 15, pp. 8647-8658, Apr 21 2020.

[40] J. B. Yoo *et al.*, "Titanium oxynitride microspheres with the rock-salt structure for use as visible-light photocatalysts," *Journal of Materials Chemistry A,* vol. 4, no. 3, pp. 869-876, 2016.
[41] B. Avasarala and P. Haldar, "Electrochemical oxidation behavior of titanium nitride based electrocatalysts under PEM fuel cell conditions," *Electrochimica Acta,* vol. 55, no. 28, pp. 9024-9034, 2010.
[42] M.-H. Chan and F.-H. Lu, "X-ray photoelectron spectroscopy analyses of titanium oxynitride films prepared by magnetron sputtering using air/Ar mixtures," *Thin Solid Films,* vol. 517, no. 17, pp. 5006-5009, 2009.
[43] J. Narayan and B. C. Larson, "Domain epitaxy: A unified paradigm for thin film growth," *Journal of Applied Physics,* vol. 93, no. 1, pp. 278-285, 2003.
[44] A. Moatti, R. Bayati, and J. Narayan, "Epitaxial growth of rutile TiO2 thin films by oxidation of TiN/Si{100} heterostructure," *Acta Materialia,* vol. 103, pp. 502-511, 2016.
[45] X. Tian *et al.*, "Transition Metal Nitride Coated with Atomic Layers of Pt as a Low-Cost, Highly Stable Electrocatalyst for the Oxygen Reduction Reaction," *J Am Chem Soc,* vol. 138, no. 5, pp. 1575-83, Feb 10 2016.
[46] P. Patsalas, N. Kalfagiannis, and S. Kassavetis, "Optical properties and plasmonic performance of titanium nitride," *Materials,* vol. 8, no. 6, pp. 3128-3154, 2015.
[47] H. Reddy, U. Guler, Z. Kudyshev, A. V. Kildishev, V. M. Shalaev, and A. Boltasseva, "Temperature-dependent optical properties of plasmonic titanium nitride thin films," *ACS photonics,* vol. 4, no. 6, pp. 1413-1420, 2017.
[48] D. Shah *et al.*, "Thickness-dependent drude plasma frequency in transdimensional plasmonic TiN," *Nano Letters,* vol. 22, no. 12, pp. 4622-4629, 2022.
[49] D. Shah, H. Reddy, N. Kinsey, V. M. Shalaev, and A. Boltasseva, "Optical properties of plasmonic ultrathin TiN films," *Advanced Optical Materials,* vol. 5, no. 13, p. 1700065, 2017.
[50] D. Shah *et al.*, "Controlling the plasmonic properties of ultrathin TiN films at the atomic level," *Acs Photonics,* vol. 5, no. 7, pp. 2816-2824, 2018.
[51] Q. Hao *et al.*, "VO2/TiN plasmonic thermochromic smart coatings for room-temperature applications," *Advanced materials,* vol. 30, no. 10, p. 1705421, 2018.
[52] A. Alvarez Barragan, N. V. Ilawe, L. Zhong, B. M. Wong, and L. Mangolini, "A non-thermal plasma route to plasmonic TiN nanoparticles," *The Journal of Physical Chemistry C,* vol. 121, no. 4, pp. 2316-2322, 2017.
[53] W. D. R. Callister, David G., *The Structure of Crystalline Solids Materials Science and Engineering an Introduction*, 10 ed. John Wiley & Sons, Inc, 2007.
[54] D. Rasic, R. Sachan, M. F. Chisholm, J. Prater, and J. Narayan, "Room Temperature Growth of Epitaxial Titanium Nitride Films by Pulsed Laser Deposition," *Crystal Growth & Design,* vol. 17, no. 12, pp. 6634-6640, 2017.
[55] J. Narayan, "Recent progress in thin film epitaxy across the misfit scale (2011 Acta Gold Medal Paper)," *Acta Materialia,* vol. 61, no. 8, pp. 2703-2724, 2013.
[56] (2024, 07/12/24). *Material Project*. Available: https://next-gen.materialsproject.org/materials/mp-1143?formula=Al2O3
[57] K. Momma and F. Izumi, "VESTA 3for three-dimensional visualization of crystal, volumetric and morphology data," *Journal of Applied Crystallography,* vol. 44, no. 6, pp. 1272-1276, 2011.
[58] L. Xiao and W. F. Schneider, "Surface termination effects on metal atom adsorption on α-alumina," *Surface Science,* vol. 602, no. 21, pp. 3445-3453, 2008.
[59] Y. Yue *et al.*, "Structure and reactivity of α-Al2O3 (0001) surfaces: how do Al–I and gibbsite-like terminations interconvert?," *The Journal of Physical Chemistry C,* vol. 126, no. 31, pp. 13467-13476, 2022.
[60] M. Roy *et al.*, "Quantum interference effects in titanium nitride films at low temperatures," *Thin Solid Films,* vol. 681, pp. 1-5, 2019.

[61] S. Gupta, A. Moatti, A. Bhaumik, R. Sachan, and J. Narayan, "Room-temperature ferromagnetism in epitaxial titanium nitride thin films," *Acta Materialia,* vol. 166, pp. 221-230, 2019.

[62] D. Jaeger and J. Patscheider, "A complete and self-consistent evaluation of XPS spectra of TiN," *Journal of Electron Spectroscopy and Related Phenomena,* vol. 185, no. 11, pp. 523-534, 2012.

[63] G. Greczynski and L. Hultman, "A step-by-step guide to perform x-ray photoelectron spectroscopy," *Journal of Applied Physics,* vol. 132, no. 1, 2022.

[64] D. N. G. Krishna and J. Philip, "Review on surface-characterization applications of X-ray photoelectron spectroscopy (XPS): Recent developments and challenges," *Applied Surface Science Advances,* vol. 12, 2022.

[65] P. M. Korusenko *et al.*, "The structure of composite coatings based on titanium nitride, formed using condensation with ion bombardment," *Journal of Physics: Conference Series,* vol. 1441, no. 1, 2020.

[66] J. Graciani, J. Fdez Sanz, T. Asaki, K. Nakamura, and J. A. Rodriguez, "Interaction of oxygen with TiN(001): N<-->O exchange and oxidation process," *J Chem Phys,* vol. 126, no. 24, p. 244713, Jun 28 2007.

[67] C. Gebauer *et al.*, "Performance of titanium oxynitrides in the electrocatalytic oxygen evolution reaction," *Nano Energy,* vol. 29, pp. 136-148, 2016.

[68] M. C. Biesinger, L. W. M. Lau, A. R. Gerson, and R. S. C. Smart, "Resolving surface chemical states in XPS analysis of first row transition metals, oxides and hydroxides: Sc, Ti, V, Cu and Zn," *Applied Surface Science,* vol. 257, no. 3, pp. 887-898, 2010.

[69] M. Maarouf, M. B. Haider, Q. A. Drmosh, and M. B. Mekki, "X-Ray Photoelectron Spectroscopy Depth Profiling of As-Grown and Annealed Titanium Nitride Thin Films," *Crystals,* vol. 11, no. 3, 2021.

[70] N. X.-r. P. S. Database, "NIST X-ray Photoelectron Spectroscopy Database Database Number 20," *National Institute of Standards and Technology,* 2000.

[71] T. F. Scientific. (2017, 4/30/24 ). *XPS Avantage Knowledge*. Available: https://www.thermofisher.com/us/en/home/materials-science/learning-center/periodic-table/transition-metal

[72] A. R. Zanatta, F. Cemin, F. G. Echeverrigaray, and F. Alvarez, "On the relationship between the Raman scattering features and the Ti-related chemical states of TixOyNz films," *Journal of Materials Research and Technology,* vol. 14, pp. 864-870, 2021.

[73] S. Oktay, Z. Kahraman, M. Urgen, and K. Kazmanli, "XPS investigations of tribolayers formed on TiN and (Ti, Re) N coatings," *Applied Surface Science,* vol. 328, pp. 255-261, 2015.

[74] B. R. KC, D. Kumar, and B. P. Bastakoti, "Block copolymer-mediated synthesis of TiO2/RuO2 nanocomposite for efficient oxygen evolution reaction," *Journal of Materials Science,* pp. 1-14, 2024.

[75] L. Soriano *et al.*, "Chemical changes induced by sputtering in TiO2 and some selected titanates as observed by X-ray absorption spectroscopy," *Surface science,* vol. 290, no. 3, pp. 427-435, 1993.

[76] Y. Hu *et al.*, "A study of titanium nitride diffusion barriers between aluminium and silicon by X-ray absorption spectroscopy: the Si, Ti and N results," *Journal of Synchrotron Radiation,* vol. 8, no. 2, pp. 860-862, 2001.

[77] V. Lusvardi, M. Barteau, J. G. Chen, J. Eng Jr, B. Frühberger, and A. Teplyakov, "An NEXAFS investigation of the reduction and reoxidation of TiO2 (001)," *Surface Science,* vol. 397, no. 1-3, pp. 237-250, 1998.

[78] F. Esaka *et al.*, "Comparison of surface oxidation of titanium nitride and chromium nitride films studied by x-ray absorption and photoelectron spectroscopy," *Journal of Vacuum Science & Technology A: Vacuum, Surfaces, and Films,* vol. 15, no. 5, pp. 2521-2528, 1997.

[79] F. Frati, M. O. Hunault, and F. M. De Groot, "Oxygen K-edge X-ray absorption spectra," *Chemical reviews,* vol. 120, no. 9, pp. 4056-4110, 2020.

[80] R. Athle, A. E. Persson, A. Irish, H. Menon, R. Timm, and M. Borg, "Effects of tin top electrode texturing on ferroelectricity in hf1–x zr x o2," *ACS applied materials & interfaces,* vol. 13, no. 9, pp. 11089-11095, 2021.
[81] S. Prayakarao et al., "Gyroidal titanium nitride as nonmetallic metamaterial," *Optical Materials Express,* vol. 5, no. 6, 2015.
[82] R. Mishra et al., "Optimized Titanium Nitride Epitaxial Film for Refractory Plasmonics and Solar Energy Harvesting," *The Journal of Physical Chemistry C,* vol. 125, no. 24, pp. 13658-13665, 2021.
[83] X.-H. Gao, Z.-M. Guo, Q.-F. Geng, P.-J. Ma, A.-Q. Wang, and G. Liu, "Enhanced optical properties of TiN-based spectrally selective solar absorbers deposited at a high substrate temperature," *Solar Energy Materials and Solar Cells,* vol. 163, pp. 91-97, 2017.
[84] R. P. Sugavaneshwar, S. Ishii, T. D. Dao, A. Ohi, T. Nabatame, and T. Nagao, "Fabrication of Highly Metallic TiN Films by Pulsed Laser Deposition Method for Plasmonic Applications," *ACS Photonics,* vol. 5, no. 3, pp. 814-819, 2017.
[85] D. Shah, H. Reddy, N. Kinsey, V. M. Shalaev, and A. Boltasseva, "Optical Properties of Plasmonic Ultrathin TiN Films," *Advanced Optical Materials,* vol. 5, no. 13, 2017.
[86] A. F. Lipinski et al., "Synthesis of Plasmonically Active Titanium Nitride Using a Metallic Alloy Buffer Layer Strategy," *ACS Appl Electron Mater,* vol. 5, no. 12, pp. 6929-6937, Dec 26 2023.
[87] B. N. Gunaydin et al., "Plasmonic Titanium Nitride Nanohole Arrays for Refractometric Sensing," *ACS Appl Nano Mater,* vol. 6, no. 22, pp. 20612-20622, Nov 24 2023.
[88] M. Popovic, M. Novakovic, and N. Bibic, "Annealing effects on the properties of tin thin films," *Processing and Application of Ceramics,* vol. 9, no. 2, pp. 67-71, 2015.
[89] O. D. Iakobson, O. L. Gribkova, A. R. Tameev, and J. M. Nunzi, "A common optical approach to thickness optimization in polymer and perovskite solar cells," *Sci Rep,* vol. 11, no. 1, p. 5005, Mar 2 2021.
[90] R. J. Xie, H. T. Hintzen, and D. Johnson, "Optical Properties of (Oxy)Nitride Materials: A Review," *Journal of the American Ceramic Society,* vol. 96, no. 3, pp. 665-687, 2013.
[91] W.-P. Guo et al., "Titanium Nitride Epitaxial Films as a Plasmonic Material Platform: Alternative to Gold," *ACS Photonics,* vol. 6, no. 8, pp. 1848-1854, 2019.
[92] a. P. C. Tanner D. (2024, 6/30). *"Datan," Data analysis package*.
[93] R. S. Krishnan, "The Raman spectrum of rocksalt and its interpretation," *Proceedings of the Indian Academy of Sciences - Section A,* vol. 26, no. 6, 1947.
[94] P. B. Johnson and R.-W. Christy, "Optical constants of the noble metals," *Physical review B,* vol. 6, no. 12, p. 4370, 1972.
[95] R. Nawaz, C. F. Kait, H. Y. Chia, M. H. Isa, and L. W. Huei, "Glycerol-Mediated Facile Synthesis of Colored Titania Nanoparticles for Visible Light Photodegradation of Phenolic Compounds," *Nanomaterials (Basel),* vol. 9, no. 11, Nov 8 2019.
[96] M. Lubas, J. J. Jasinski, M. Sitarz, L. Kurpaska, P. Podsiad, and J. Jasinski, "Raman spectroscopy of TiO2 thin films formed by hybrid treatment for biomedical applications," *Spectrochim Acta A Mol Biomol Spectrosc,* vol. 133, pp. 867-71, Dec 10 2014.
[97] E. Perevedentseva et al., "Raman Spectroscopic Study of TiO(2) Nanoparticles' Effects on the Hemoglobin State in Individual Red Blood Cells," *Materials (Basel),* vol. 14, no. 20, Oct 9 2021.
[98] S. Sahoo, A. K. Arora, and V. Sridharan, "Raman Line Shapes of Optical Phonons of Different Symmetries in Anatase TiO2 Nanocrystals," *The Journal of Physical Chemistry C,* vol. 113, no. 39, pp. 16927-16933, 2009.
[99] S. Challagulla, K. Tarafder, R. Ganesan, and S. Roy, "Structure sensitive photocatalytic reduction of nitroarenes over TiO (2)," *Sci Rep,* vol. 7, no. 1, p. 8783, Aug 18 2017.
[100] W. Spengler, R. Kaiser, A. N. Christensen, and G. Müller-Vogt, "Raman scattering, superconductivity, and phonon density of states of stoichiometric and nonstoichiometric TiN," *Physical Review B,* vol. 17, no. 3, pp. 1095-1101, 1978.

[101] Y. H. Cheng, B. K. Tay, S. P. Lau, H. Kupfer, and F. Richter, "Substrate bias dependence of Raman spectra for TiN films deposited by filtered cathodic vacuum arc," *Journal of Applied Physics,* vol. 92, no. 4, pp. 1845-1849, 2002.

[102] C. Moura, P. Carvalho, F. Vaz, L. Cunha, and E. Alves, "Raman spectra and structural analysis in ZrOxNy thin films," *Thin Solid Films,* vol. 515, no. 3, pp. 1132-1137, 2006.

[103] A. N. D. Christensen, O. W.; Kress, W.; Teuchert, W. D, "Phonon anomalies in transition-metal nitrides: ZrN," *Phys. Rev. B,* vol. 19, no. 11, pp. 5699--5703, 1979. American Physical Society

[104] A. Cassinese, M. Iavarone, R. Vaglio, M. Grimsditch, and S. Uran, "Transport properties of ZrN superconducting films," *Physical Review B,* vol. 62, no. 21, pp. 13915-13918, 2000.

[105] I. Burducea *et al.*, "A new ion beam facility based on a 3 MV Tandetron™ at IFIN-HH, Romania," *Nuclear Instruments and Methods in Physics Research Section B: Beam Interactions with Materials and Atoms,* vol. 359, pp. 12-19, 2015.

[106] M. Mayer, "SIMNRA User's Guide. Report IPP 9/113," *Max-Planck-Institut für Plasmaphysik, Garching, Germany,* 1997.

[107] G. Velişa *et al.*, "Joint research activities at the 3 MV Tandetron™ from IFIN-HH," *The European Physical Journal Plus,* vol. 136, no. 11, 2021.

Supplementary information

# Optical and plasmonic properties of high electron density epitaxial and oxidative controlled titanium nitride thin films

Ikenna Chris-Okoro[1], Sheilah Cherono[1], Wisdom Akande[1], Swapnil Nalawade[2], Mengxin Liu[1], Catalin Martin[3], Valentin Craciun[1,4], R. Soyoung Kim[5], Johannes Mahl[5], Tanja Cuk[6], Junko Yano[5], Ethan Crumlin[5], J. David Schall[1], Shyam Aravamudhan[2], Maria Diana Mihai[7,8], Jiongzhi Zheng[9], Lei Zhang[9], Geoffroy Hautier[9], and Dhananjay Kumar[1*]

[1]Department of Mechanical Engineering, North Carolina A&T State University, Greensboro, NC 27411, USA

[2]Joint School of Nanoscience and Nanoengineering, North Carolina A&T State University, Greensboro, NC, USA 27401

[3]School of Theoretical & Applied Sciences, Ramapo College of New Jersey, Mahwah, NJ 07430, USA

[4]National Institute for Laser, Plasma, and Radiation Physics and Extreme Light Infrastructure for Nuclear Physics, RO 060042, Magurele, Romania

[5]Lawrence Berkeley National Laboratory, Berkeley, CA 94720

[6]Department of Chemistry, University of Colorado, Boulders, CO 80309, USA

[7]Horia Hulubei National Institute for Physics and Nuclear Engineering, Măgurele, IF, 077125, Romania

[8]Department of Physics, National University of Science and Technology Politehnica Bucharest, RO, 060042, Romania

[9]Department of Materials Science and Engineering, Dartmouth College, NH, USA 03755

*Email: dkumar@ncat.edu

**Table of Contents**

**List of Tables**

**Table of Figure**

**Table of Note**

|  | Resistivity (μΩcm) | XRD Peak Index | Peak Position 2θ(°) | FWHM β(°) | Thickness XRR (nm) | Density (gcm$^{-3}$) |
|---|---|---|---|---|---|---|
| **700°C, Vacuum** | 190 | (111) | 36.732 | 0.375 | 300 | 5.430 (4.47% acc.) |
|  |  | (222) | 78.256 | 0.599 |  |  |
| **600°C, Vacuum** | 407 | (111) | 37.136 | 0.257 | 312 | 5.327 (4.25% acc.) |
|  |  | (222) | 79.098 | 0.388 |  |  |
| **700°C, 5 mTorr O$_2$** | 365 | (111) | 37.009 | 0.315 | 302 | 5.019 (4.37% acc.) |
|  |  | (222) | 78.892 | 0.562 |  |  |

**Table S1 XRD & XRR Analysis**. Calculated values for resistivity, Peak Index, XRD Peak Broadening, Thickness and dislocation density for TiN$_x$O$_y$ films

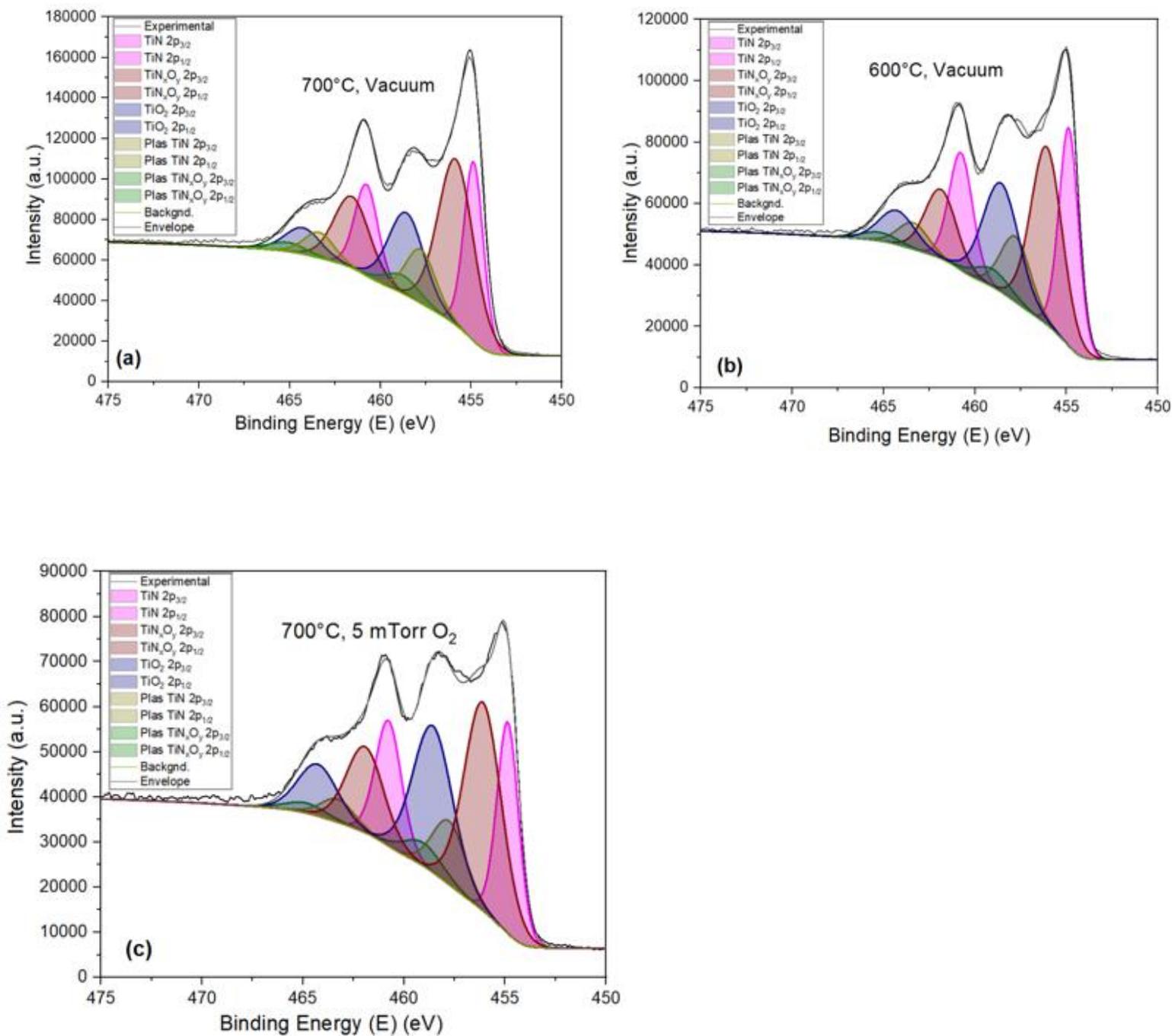

**Figure S1 Ti 2p deconvoluted XPS Spectra** (a) 700°C, Vacuum (b) 600°C, Vacuum (c) 700°C, 5 mTorr O$_2$. As it is well reported in literature, the Ti 2p 3/2 is expected to be two times the relative area of the Ti 2p 1/2. However, due to Coster-Kronig effects[1-4], this may not always be so

|  |  | Binding Energies | | | FWHM | | |
|---|---|---|---|---|---|---|---|
|  |  | 700°C, Vacuum | 600°C, Vacuum | 700°C, 5 mTorr $O_2$ | 700°C, Vacuum | 600°C, Vacuum | 700°C, 5 mTorr $O_2$ |
| TiN | 2p $_{3/2}$ | 454.87 | 454.87 | 454.87 | 1.31 | 1.31 | 1.31 |
|  | 2p $_{1/2}$ | 460.77 | 460.77 | 460.77 | 1.57 | 1.57 | 1.57 |
| TiN$_x$O$_y$ | 2p $_{3/2}$ | 455.85 | 456.09 | 455.85 | 2.40 | 2.09 | 2.40 |
|  | 2p $_{1/2}$ | 461.57 | 461.87 | 461.57 | 2.40 | 2.09 | 2.40 |
| TiO$_2$ | 2p $_{3/2}$ | 458.58 | 458.58 | 458.58 | 2.40 | 2.40 | 2.40 |
|  | 2p $_{1/2}$ | 464.28 | 464.28 | 464.28 | 2.40 | 2.40 | 2.40 |
| Plas TiN | 2p $_{3/2}$ | 457.78 | 457.78 | 457.78 | 2.00 | 2.00 | 2.00 |
|  | 2p $_{1/2}$ | 463.38 | 463.38 | 463.38 | 2.00 | 2.00 | 2.00 |
| Plas TiN$_x$O$_y$ | 2p $_{3/2}$ | 458.80 | 459.04 | 458.80 | 2.40 | 2.40 | 2.40 |
|  | 2p $_{1/2}$ | 465.10 | 465.24 | 465.10 | 2.40 | 2.40 | 2.40 |
| TiN Diff | | 5.90 | 5.90 | 5.90 | | | |
| TiNO Diff | | 5.72 | 5.78 | 5.82 | | | |
| TiO$_2$ Diff | | 5.70 | 5.70 | 5.70 | | | |

**Table S2 Ti 2p fitting Parameter**. Binding Energies and FWHM of the various species of the decovoluted Ti 2p Spectra as well as the binding energy difference from the 2 halves.

| % Relative Molar Fract. | 700°C, Vacuum | 600°C, Vacuum | 700°C, 5 mTorr $O_2$ |
|---|---|---|---|
| TiN (Ti 2p) | 28.22 | 31.08 | 25.25 |
| $TiN_xO_y$ (Ti 2p) | 47.94 | 40.49 | 42.79 |
| $TiO_2$ (Ti 2p) | 23.84 | 28.43 | 31.96 |
| Ti-N (N 1s) | 33.67 | 38.77 | 29.63 |
| Ti-N-O (N 1s) | 53.85 | 46.46 | 57.16 |
| N-O (N 1s) | 12.48 | 14.77 | 13.21 |
| Ti-O (O 1s) | 23.35 | 31.73 | 32.52 |
| Ti-N-O (O 1s) | 45.08 | 45.1 | 44.03 |
| Ti-Carbonate/Hydroxyl (O1s) | 22.14 | 16.19 | 17.03 |
| Absorbed O (OH-C=O) (O 1s) | 9.48 | 6.98 | 6.42 |

**Table S3 Relative Molar Fraction from the Ti 2p, O 1s and N 1s XPS Spectra**. Molar fraction of these various species present in the film deconvoluted from the Ti 2p, O 1s and N 1s XPS Spectra

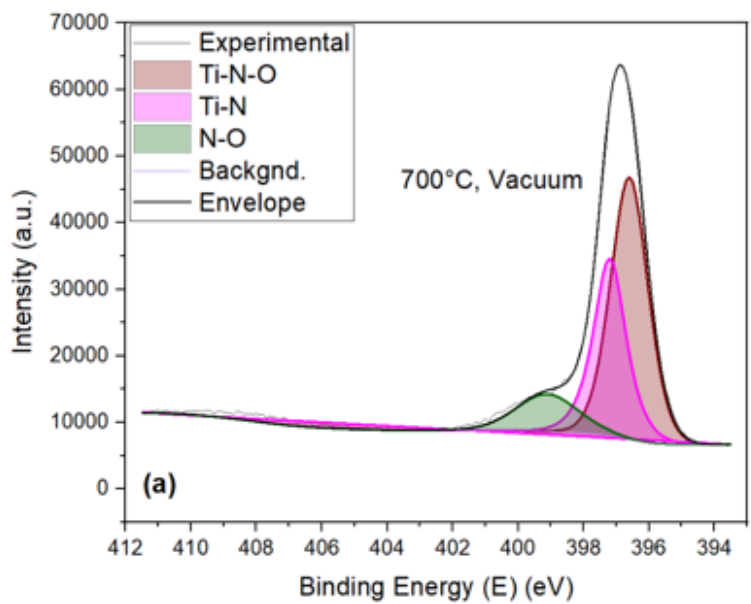
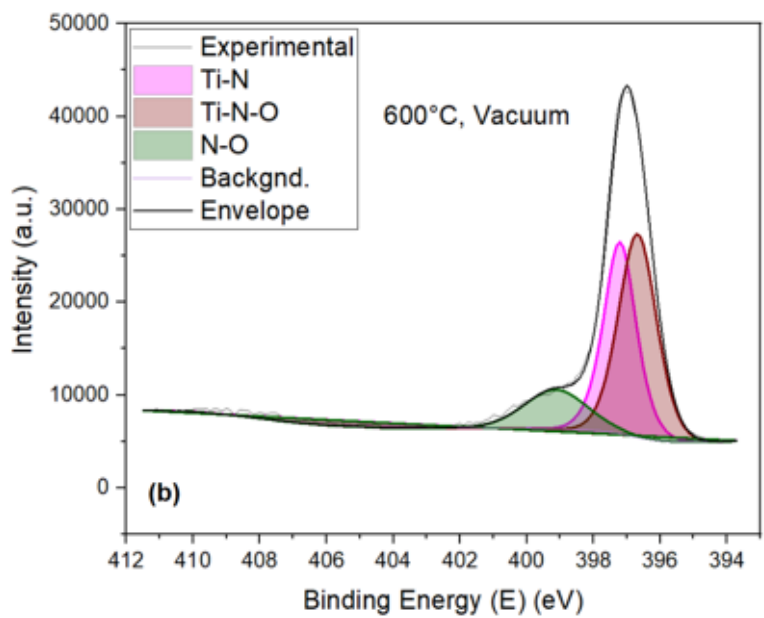
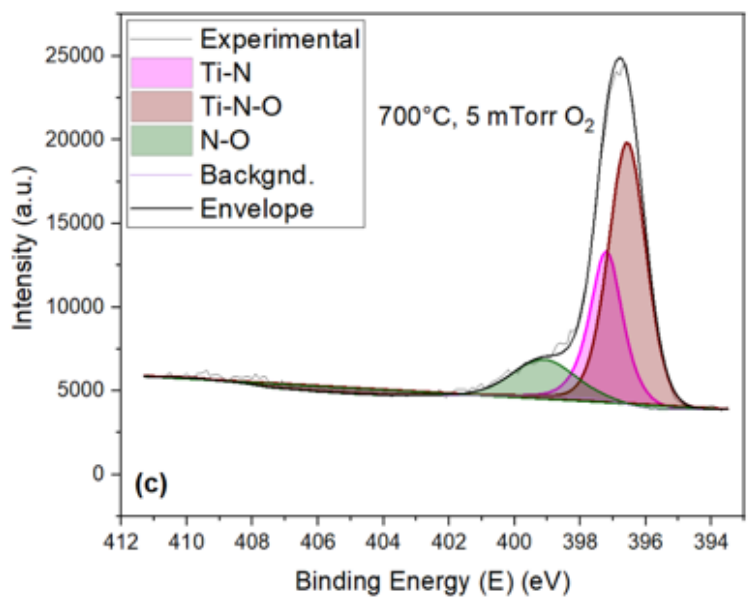

**Figure S2 N 1s deconvoluted XPS Spectra** for (a) 700°C, Vacuum (b) 600°C, Vacuum (c) 700°C, 5 mTorr $O_2$

|        | Binding Energies |                |                            | FWHM          |                |                            |
|--------|------------------|----------------|----------------------------|---------------|----------------|----------------------------|
|        | 700°C, Vacuum    | 600°C, Vacuum  | 700°C, 5 mTorr $O_2$       | 700°C, Vacuum | 600°C, Vacuum  | 700°C, 5 mTorr $O_2$       |
| **Ti-N**   | 397.20           | 397.20         | 397.20                     | 1.20          | 1.20           | 1.20                       |
| **Ti-N-O** | 396.60           | 396.66         | 396.68                     | 1.40          | 1.40           | 1.40                       |
| **N-O**    | 399.13           | 399.13         | 399.13                     | 2.30          | 2.30           | 2.30                       |

**Table S4 N 1s fitting Parameter.** Binding Energies and FWHM of the various species of the decovoluted N 1s Spectra and the molar fraction of these species.

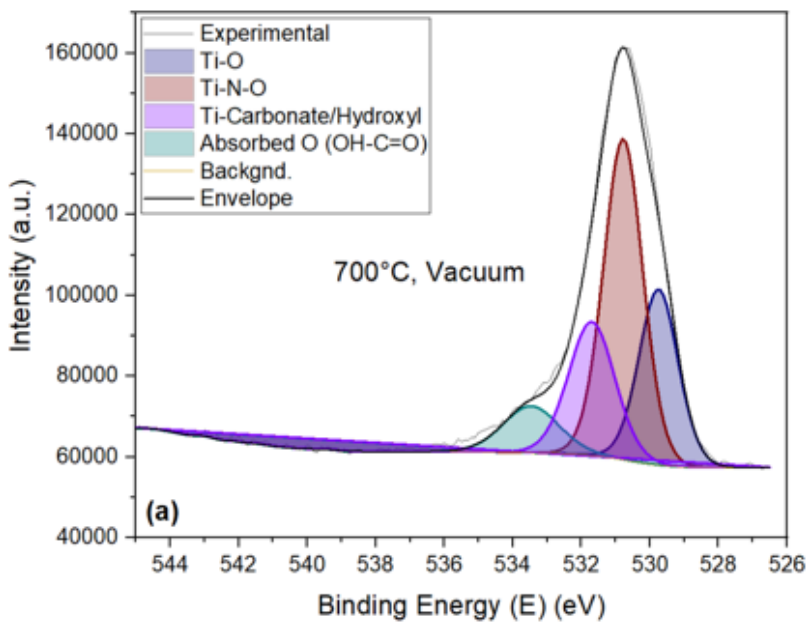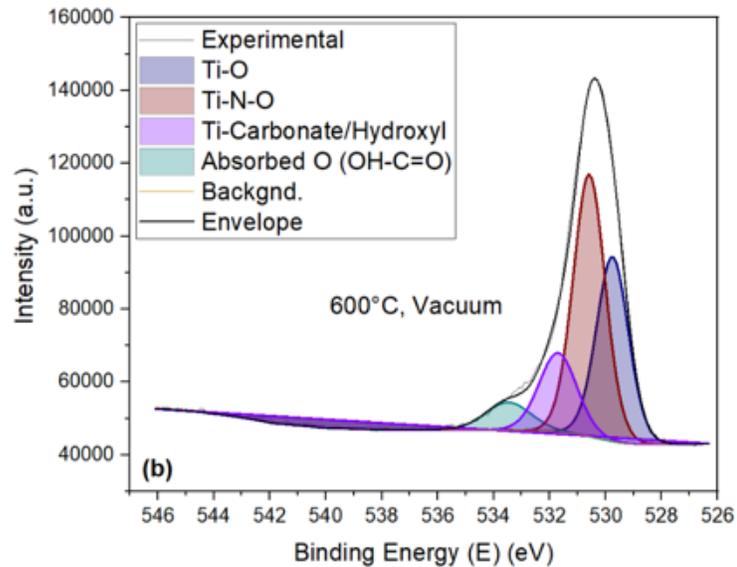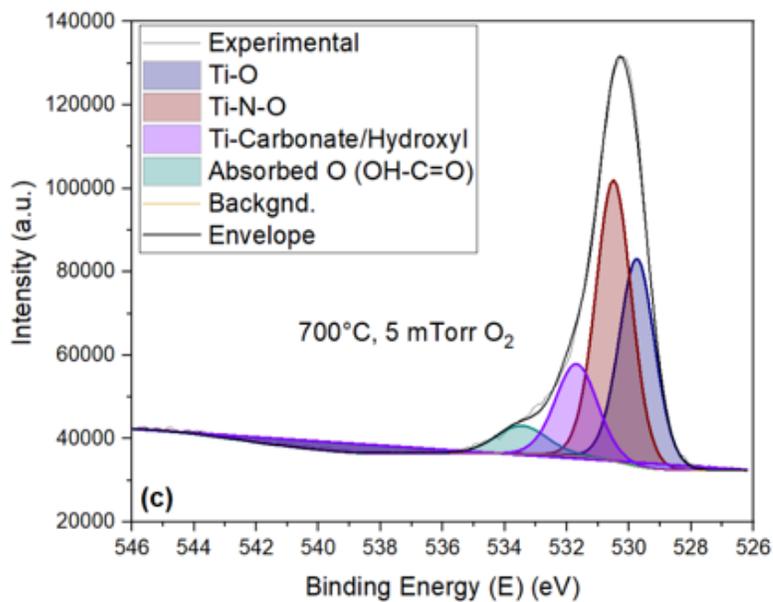

**Figure S3 O 1s deconvoluted XPS Spectra** for (a) 700°C, Vacuum (b) 600°C, Vacuum (c) 700°C, 5 mTorr $O_2$

|  | Binding Energies | | | FWHM | | |
|---|---|---|---|---|---|---|
|  | 700°C, Vacuum | 600°C, Vacuum | 700°C, 5 mTorr $O_2$ | 700°C, Vacuum | 600°C, Vacuum | 700°C, 5 mTorr $O_2$ |
| **Ti-O** | 529.75 | 529.75 | 529.75 | 1.40 | 1.40 | 1.40 |
| **Ti-N-O** | 530.78 | 530.58 | 530.50 | 1.40 | 1.40 | 1.40 |
| **Ti-Carbonate/Hydroxyl** | 531.70 | 531.70 | 531.70 | 1.65 | 1.65 | 1.65 |
| **Absorbed O (OH-C=O)** | 533.47 | 533.47 | 533.47 | 2.00 | 2.00 | 2.00 |

**Table S4 O 1s fitting Parameter.** Binding Energies and FWHM of the various species of the decovoluted O 1s Spectra.

| Conditions | Elemental Percent | | | Chemical Formula | | $Ti^{3+}$ vacancy per unit cell | $Ti^{3+}$ defect density per $cm^3$ | |
| --- | --- | --- | --- | --- | --- | --- | --- | --- |
| | Ti | N | O | Actual | Equivalent (showing excess O, $O_{Exc}$) | | Total | Actual TiNO |
| 700°C, Vacuum | 39.42 | 22.66 | 37.92 | $TiN_{0.57}O_{0.96}$ | $TiN_{0.57}O_{0.43+0.53}$ | 0.14 | 1.84E+21 | 8.84E+20 |
| 600°C, Vacuum | 39.10 | 20.13 | 40.77 | $TiN_{0.51}O_{1.04}$ | $TiN_{0.51}O_{0.49+0.55}$ | 0.16 | 2.18E+21 | 8.81E+20 |
| 700°C, 5 mTorr $O_2$ | 37.77 | 14.03 | 48.20 | $TiN_{0.37}O_{1.28}$ | $TiN_{0.37}O_{0.63+0.65}$ | 0.21 | 2.82E+21 | 1.21E+21 |

**Table S6 XPS Elemental and Structural composition**, $Ti^{3+}$ Defect Density Calculation from the TiNO homolog Series Formula $Ti^{3+}_{1-x/3}Ti(vac)^{3+}_{x/3}N^{3-}_{1-x}O^{2-}_{x}$ of the $TiN_xO_y$ films from XPS Results.

| Sample name | Characterization |||||
| --- | --- | --- | --- | --- | --- |
| | Layers | Thickness (nm) | Composition |||
| Nitrogen Resonance 700°C, Vacuum | 1 | 300 nm $TiN_{0.66}O_{0.34}$ | N 0.33 | Ti 0.50 | O 0.17 |
| | 2* | 500000 | Al 0.40 || O 0.60 |
| Nitrogen Resonance 600°C, Vacuum | 1 | 314 nm $TiN_{0.63}O_{0.69}$ | N 0.27 | Ti 0.43 | O 0.30 |
| | 2* | 500000 | Al 0.40 || O 0.60 |
| Nitrogen Resonance 700°C, 5 mTorr $O_2$ | 1 | 302 nm $TiN_{0.46}O_{0.99}$ | N 0.19 | Ti 0.41 | O 0.40 |
| | 2* | 500000 | O 0.60 || Al 0.40 |

**Table S7 NRBS Data (Nitrogen Resonance)**. The stoichiometry and thickness, of the $TiN_xO_y$ films from the non-RBS measurements. These results as seen in **Table S5**, **Fig. S4**, and **Fig. S5** in comparison with the XPS at surface confirm the presence of various chemistries across the thickness of the film from a more oxygenated surface to a more nitrogenated bulk

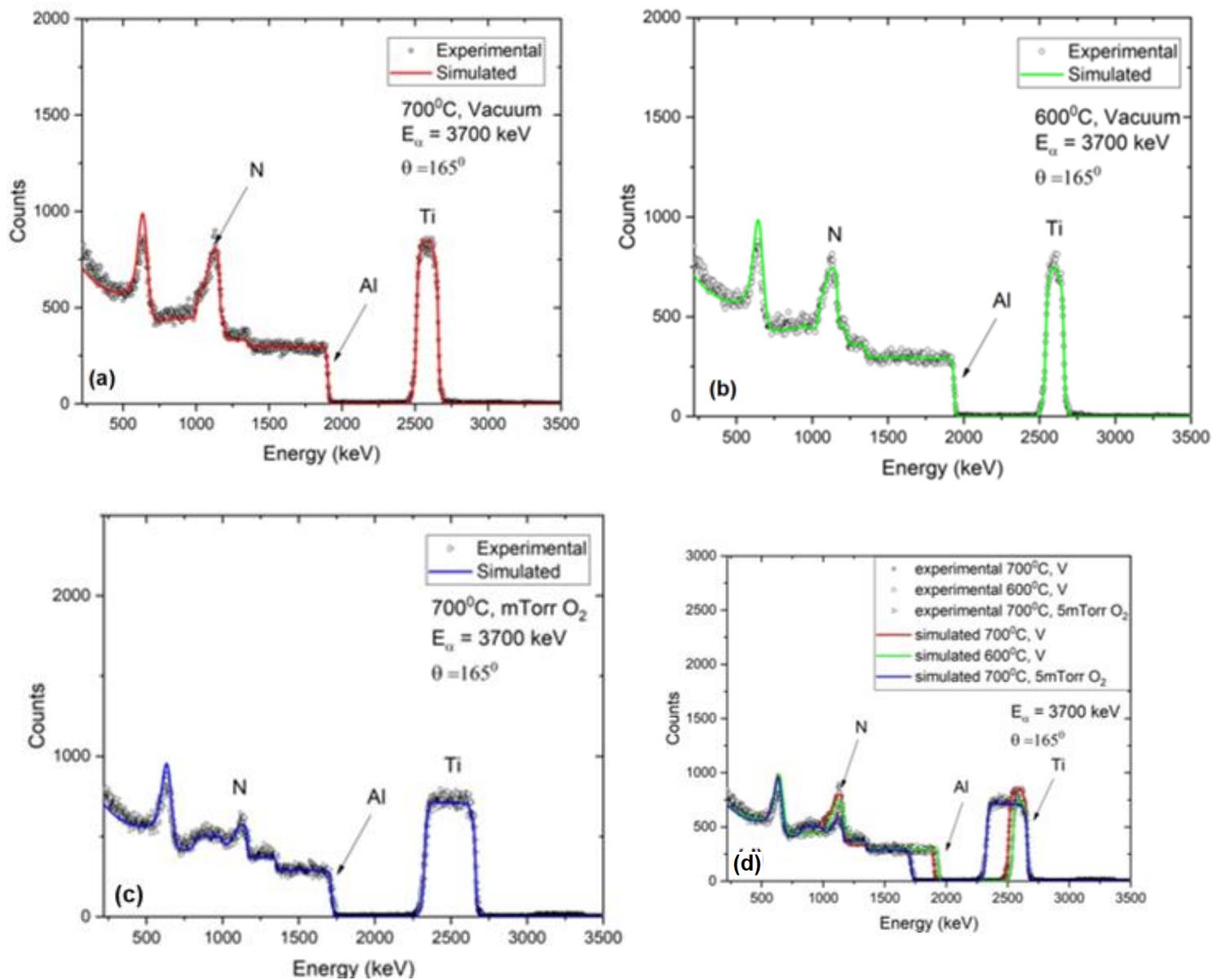

**Figure S4 Nitrogen Resonance Spectra from Non-Rutherford Backscaterring Spectrometry and fitting Simulation** (a) 700°C, Vacuum (b) 600°C, Vacuum (c) 700°C, 5 mTorr $O_2$ (d) Comparision Plot.

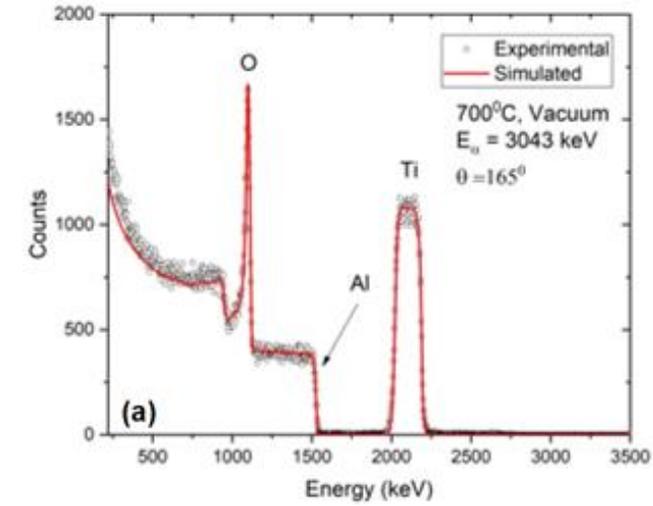
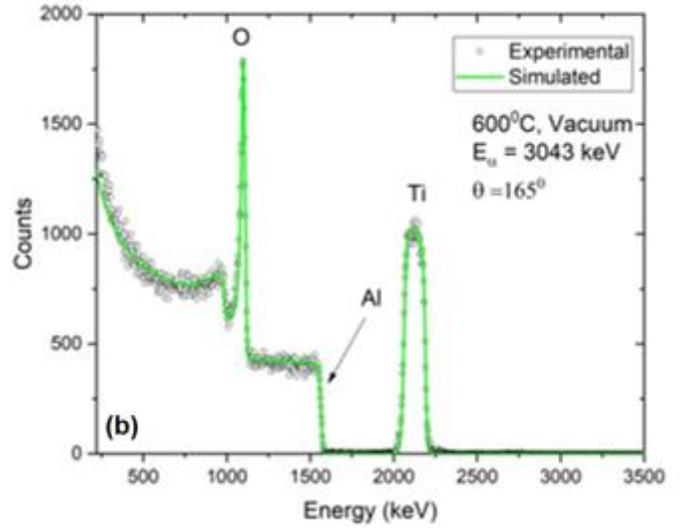
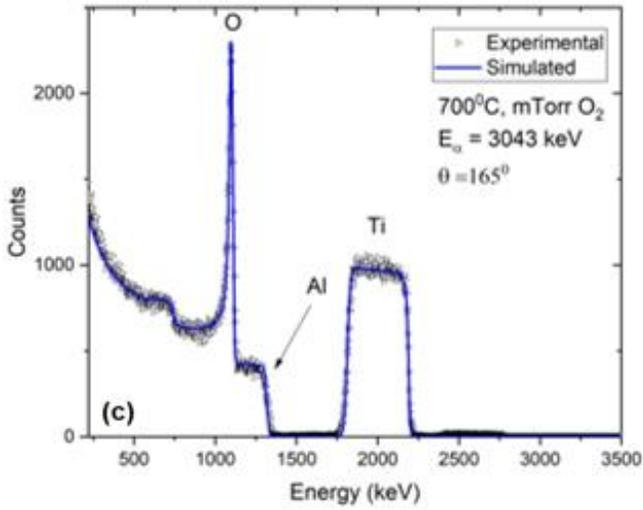
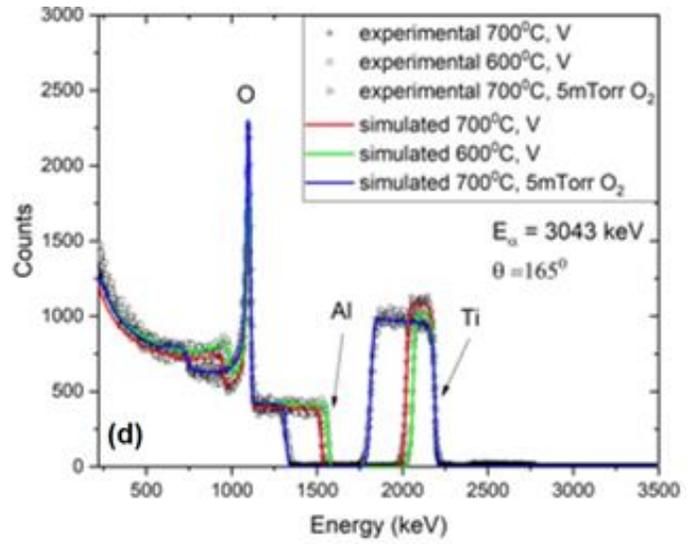

**Figure S5 Oxygen Resonance Spectra from Non-Rutherford Backscaterring Spectrometry and fitting Simulation** (a) 700°C, vacuum (b) 600°C, vacuum (c) 700°C, 5 mTorr $O_2$ (d) Comparision Plot.

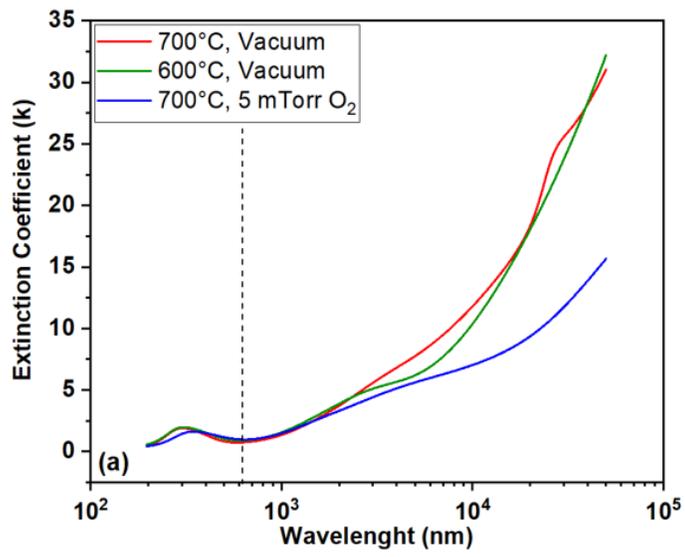 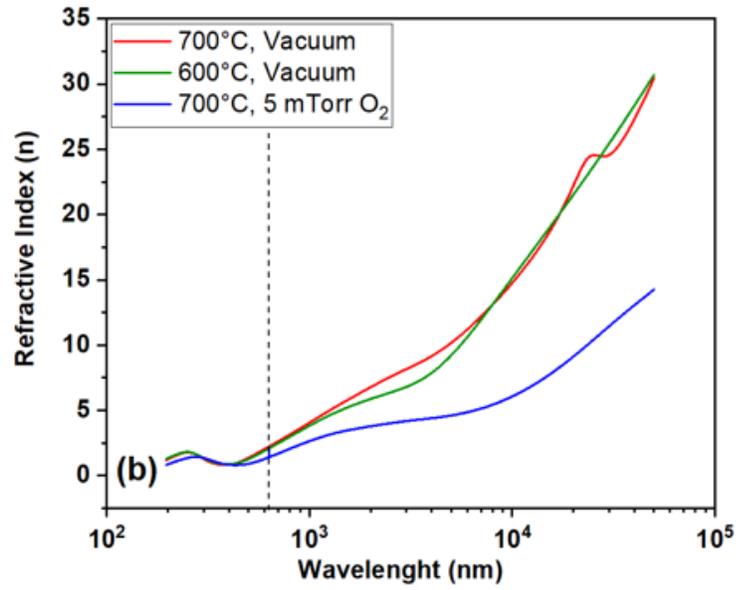

**Figure S6. Extinction Coefficient and Refractive Index**. (a) Extinction Coefficient (k). (b) refractive index (n) of the various films.

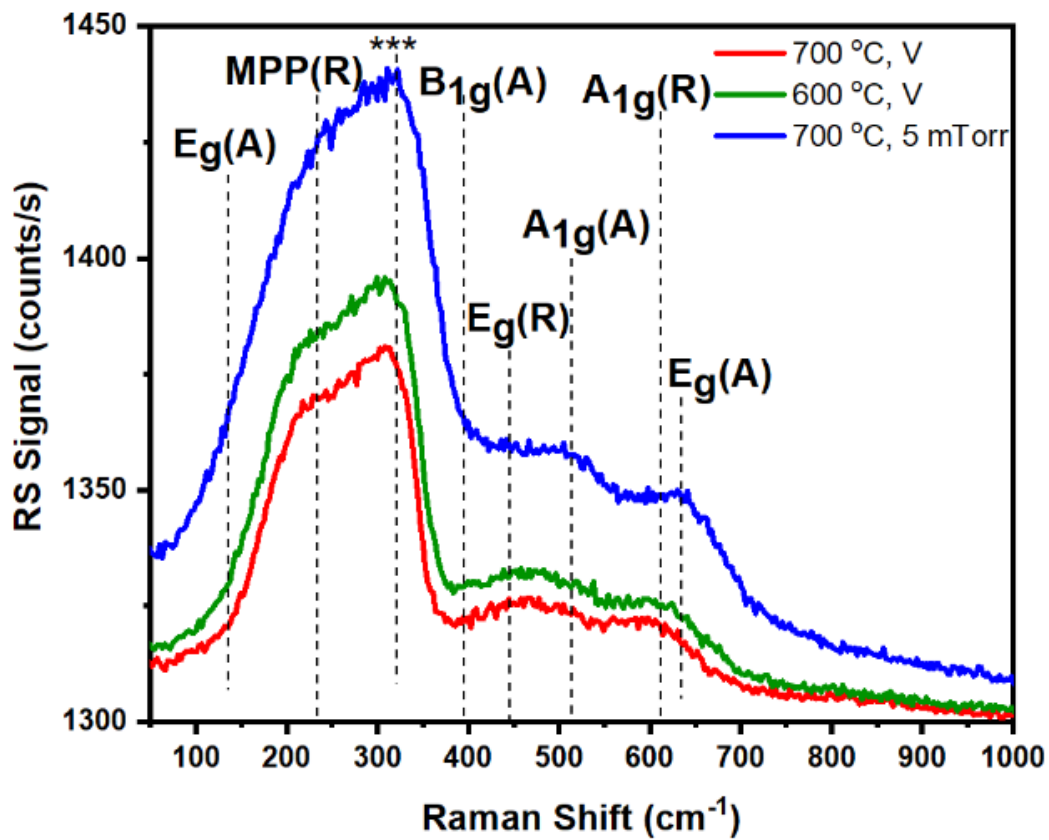

**Figure S7 Raman Spectroscopy** (a) Raman Spectra from the $TiN_xO_y$ deposited at 700°C and 600°C at vacuS1um and 5 mTorr with 532nm wavelength laser excitation. $E_g$ (144 cm$^{-1}$-A, 198 cm$^{-1}$-A, 446 cm$^{-1}$-R, 632 cm$^{-1}$-A), $B_{1g}$ (140 cm$^{-1}$-R, 394 cm$^{-1}$-A), $A_{1g}$ (514 cm$^{-1}$-R, 610 cm$^{-1}$-A), and Multi-Photon Phase-MPP (240cm$^{-1}$-R)

**Supporting Note S1**. Calculation and Discussion of the Lorentz-Drude fit model

The reflectance from Fig. 6 cannot be associated with the $Al_2O_3$ substrate as the $TiN_xO_y$ films are too thick (reference **Table S1**) to allow reflectance from the film-substrate interface. A reliability analysis based on the film -substrate thickness was undertaken across air-film, film-substrate and substrate-air interfaces in the 4 different media as seen in the inset of Fig. 8(b). with the 2 air-based media with semi-finite with a refractive index $n_0 = 1$. Applying matrix approach for the propagation of light in a medium [5-11] partially shown in the inset of Fig. 8(b) where $M_f$ and $M_s$ are transfer matrices of the film and substrate respectively with resultant phase factors $\delta_f$ and $\delta_s$, complex refractive indices $\tilde{n}_f$ and $\tilde{n}_s$ and wavelength $\lambda$ (ref eqn. 1).

$$\begin{pmatrix} E_0^+ \\ E_0^- \end{pmatrix} = M_f \cdot M_s \begin{pmatrix} E_1^+ \\ E_1^- \end{pmatrix}, \quad M_{f,s} = \begin{pmatrix} \cos \delta_{f,s} & \frac{-i}{\tilde{n}_{f,s}} \sin \delta_{f,s} \\ -i\tilde{n}_{f,s} \sin \delta_{f,s} & \cos \delta_{f,s} \end{pmatrix} \text{ and } \delta_{f,s} = 2\pi \tilde{n}_{f,s} d_{f,s}/\lambda, \quad (1)$$

If these values are known, the propagation matrix which is a product of the transfer matrices is further simplified eqn. 2, and assuming negligible amount of light reflected off the back if the substrate $E_1^- = 0$. Then, the amplitude complex reflection ($r$) and transmission ($t$) coefficients are defined as in eqn. 2.

$$M_f \cdot M_s = \begin{pmatrix} \tilde{m}_{11} & \tilde{m}_{12} \\ \tilde{m}_{21} & \tilde{m}_{22} \end{pmatrix}, \quad \tilde{r} = \frac{E_0^-}{E_0^+} = \frac{\tilde{m}_{21}}{\tilde{m}_{11}} \text{ and } \tilde{t} = \frac{E_1^+}{E_0^+} = \tilde{m}_{11} \quad (2)$$

The frequency dependence of reflectance $\mathcal{R}_{meas}(\omega)$ is analyzed assuming a Lorentz-Drude model for the complex dielectric functions of the film ($\tilde{\varepsilon}_f$) and substrate ($\tilde{\varepsilon}_s$) [1,2] as seen in eqn. 3.

$$\tilde{\varepsilon}(\omega) = \varepsilon_1(\omega) + i\varepsilon_2(\omega) = \varepsilon_\infty - \frac{\omega_p^2}{\omega^2 + \frac{i\omega}{\tau}} + \sum_{j=1}^{N} \frac{S_j^2}{\omega_j^2 - \omega^2 - \frac{i\omega}{\tau_j}} \text{ and } \tilde{n}_{f,s} = \sqrt{\tilde{\varepsilon}_{f,s}} \quad (3)$$

**Supporting Note S2.** Calculation and Discussion of the TiNO homolog series and the formation of Ti vacanies

As had been proven in our previous study [12] and utilizing the TiNO homolog series formula $Ti^{3+}_{1-x/3}Ti(vac)^{3+}_{x/3}N^{3-}_{1-x}O^{2-}_{x}$ rewritten for the sample deposited in vacuum at 700°C is $Ti^{3+}_{0.86}Ti(vac)^{3+}_{0.14}N^{3-}_{0.57}O^{2-}_{0.43+0.53}$ (ref. **Table S5**) and from the lattice calculation the unit cell is maintains its electrical neutrality as it ejects 0.14 $Ti^{3+}$ cations from each unit cell (~1 $Ti^{3+}$ cation in every 7 unit cells) which will naturally combine with the excess 0.53 $O^{2-}$ for the formation of stoichiometric $TiO_2$ ($Ti_{0.14}O_{0.28}$) and other non-stoichiometric titanium dioxide such as TiO, $Ti_2O_3$, $Ti_2O_5$. The excess oxygen is then available for the formation of carbonate, hydroxyl and with as chemisorbed oxides species as is represented in the O 1s XPS spectra. The terminal compounds (TiN with x=0 and TiO with x=1), as well as all the intermediate compounds with x between 0 and 1, all have rock salt structures (as shown in the figure below). The figure highlights the isostructural rocksalt unit cell structures of TiN, TiNO with composition studied in this work, and TiO. Also shown is the unit cell structure of rutile $TiO_2$ that is formed by a combination of Ti leaving the TiN/TiNO lattices and oxygen ambiance used during the film growth.

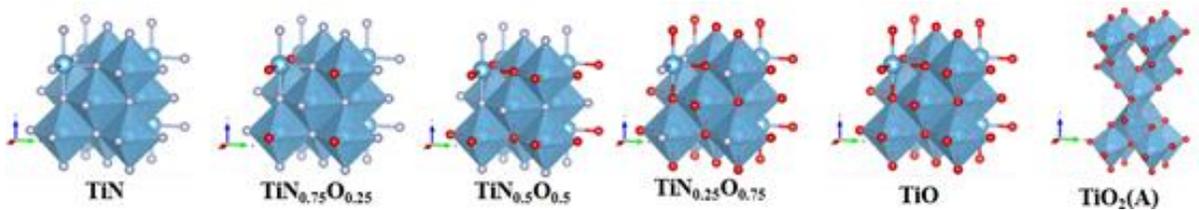

TiN    $TiN_{0.75}O_{0.25}$    $TiN_{0.5}O_{0.5}$    $TiN_{0.25}O_{0.75}$    TiO    $TiO_2(A)$

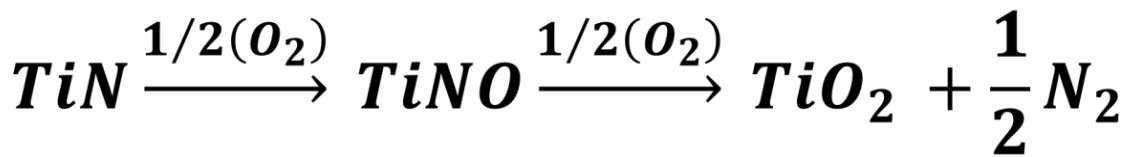

$$TiN \xrightarrow{1/2(O_2)} TiNO \xrightarrow{1/2(O_2)} TiO_2 + \frac{1}{2}N_2$$

**Supporting Note S2 Figure:** Model routes for the formation of $TiN_xO_y$ and phase transformation through oxidation of TiN, nitridation of $TiO_2$. (Vesta illustration of the ball and stick atomic model). The red, ash, and blue balls represent oxygen, nitrogen, and titanium atoms respectively. The red, green, and blue arrows indicate the a-, b-, and c-axis, respectively.

**Supporting Note S3.** First-principles calculation and lattice dynamics

In this study, ab-initio calculations based on density functional theory (DFT) [13] using the Vienna Ab-initio Simulation Package (VASP) [14] were carried out to optimize the structures of anatase and rutile TiO$_2$ crystals. The projector-augmented wave (PAW) [15]method was employed to treat the Ti($4d^{10}5s^1$) and O($2s^22p^4$) shells as valence states. The Perdew-Burke-Ernzerhof (PBE) functional[16], within the framework of the generalized gradient approximation (GGA) [17], was used for the exchange-correlation functional. The ionic positions and unit cell geometry were fully optimized using a plane-wave cutoff energy of 600 eV and a 12 × 12 × 16 Monkhorst-Pack electronic k-point mesh, with a strict force convergence criterion of $10^{-5}$ eV·Å$^{-1}$ and an energy convergence criterion of $10^{-8}$ eV. The resulting relaxed lattice constants are a = b = 4.66 Å, c = 2.97 Å for rutile, and a = b = c = 5.56 Å for anatase TiO$_2$ crystals, respectively. To account for long-range dipole-dipole interactions in the crystals, the dielectric tensor (ε) and the Born effective charges (Z) of both rutile and anatase TiO$_2$ were calculated using density functional perturbation theory (DFPT)[18]. The harmonic interatomic force constants (IFCs) were calculated using the finite-displacement approach [19] with a 3 × 3 × 3 supercell for anatase TiO$_2$ and a 3 × 3 × 4 supercell for rutile TiO$_2$, along with a 4 × 4 × 4 Monkhorst-Pack electronic k-point mesh and a plane-wave cutoff energy of 600 eV in VASP. The virtual crystal approximation (VCA) approach [20, 21]was used to model the bulk TiNO compound for lattice dynamics. In this work, the phonon calculations were performed using the Phonopy package[22].

# References

[1] Ö. Söğüt, E. Büyükkasap, M. Ertuğrul, and A. Küçükönder, "Chemical effect on enhancement of Coster–Kronig transition of L3 X-rays," *Journal of Quantitative Spectroscopy and Radiative Transfer,* vol. 74, no. 3, pp. 395-400, 2002.

[2] M. Ohno, "Effects of Coster–Kronig fluctuation and decay on X-ray photoelectron spectroscopy spectra," *Journal of electron spectroscopy and related phenomena,* vol. 131, pp. 3-28, 2003.

[3] R. Nyholm, N. Martensson, A. Lebugle, and U. Axelsson, "Auger and Coster-Kronig broadening effects in the 2p and 3p photoelectron spectra from the metals 22Ti-30Zn," *Journal of Physics F: Metal Physics,* vol. 11, no. 8, p. 1727, 1981.

[4] W. Bambynek *et al.*, "X-ray fluorescence yields, Auger, and Coster-Kronig transition probabilities," *Reviews of modern physics,* vol. 44, no. 4, p. 716, 1972.

[5] O. S. Heavens, "Optical properties of thin films," *Reports on Progress in Physics,* vol. 23, no. 1, pp. 1-65, 1960.

[6] R. Azzam and N. Bashara, "Application of generalized ellipsometry to anisotropic crystals," *JOSA,* vol. 64, no. 2, pp. 128-133, 1974.

[7] P. H. Lissberger, "Ellipsometry and polarised light," *Nature,* vol. 269, no. 5625, pp. 270-270, 1977.

[8] K. Ohta and H. Ishida, "Matrix formalism for calculation of electric field intensity of light in stratified multilayered films," *Appl Opt,* vol. 29, no. 13, pp. 1952-9, May 1 1990.

[9] D. B. Tanner, *Optical Effects in Solids*. 2019.

[10] F. Han, "Optical Properties of Solids," in *A Modern Course in the Quantum Theory of Solids*, 2012, pp. 499-579.

[11] F. Wooten, *Optical properties of solids*. Citeseer, 1972.

[12] M. Roy *et al.*, "Modulation of Structural, Electronic, and Optical Properties of Titanium Nitride Thin Films by Regulated In Situ Oxidation," *ACS Appl Mater Interfaces,* vol. 15, no. 3, pp. 4733-4742, Jan 25 2023.

[13] P. Hohenberg and W. Kohn, "Inhomogeneous electron gas," *Physical review,* vol. 136, no. 3B, p. B864, 1964.

[14] G. Kresse and J. Furthmüller, "Efficient iterative schemes for ab initio total-energy calculations using a plane-wave basis set," *Physical review B,* vol. 54, no. 16, p. 11169, 1996.

[15] P. E. Blöchl, "Projector augmented-wave method," *Physical review B,* vol. 50, no. 24, p. 17953, 1994.

[16] J. P. Perdew, K. Burke, and M. Ernzerhof, "Generalized gradient approximation made simple," *Physical review letters,* vol. 77, no. 18, p. 3865, 1996.

[17] J. P. Perdew, K. Burke, and Y. Wang, "Generalized gradient approximation for the exchange-correlation hole of a many-electron system," *Physical review B,* vol. 54, no. 23, p. 16533, 1996.

[18] S. Baroni, S. De Gironcoli, A. Dal Corso, and P. Giannozzi, "Phonons and related crystal properties from density-functional perturbation theory," *Reviews of modern Physics,* vol. 73, no. 2, p. 515, 2001.

[19] K. Esfarjani and H. T. Stokes, "Method to extract anharmonic force constants from first principles calculations," *Physical Review B—Condensed Matter and Materials Physics,* vol. 77, no. 14, p. 144112, 2008.

[20] F. Yao *et al.*, "Experimental evidence of superdiffusive thermal transport in Si0. 4Ge0. 6 thin films," *Nano Letters,* vol. 22, no. 17, pp. 6888-6894, 2022.

[21] W. Li, L. Lindsay, D. A. Broido, D. A. Stewart, and N. Mingo, "Thermal conductivity of bulk and nanowire Mg 2 Si x Sn 1− x alloys from first principles," *Physical Review B—Condensed Matter and Materials Physics,* vol. 86, no. 17, p. 174307, 2012.

[22]  A. Togo and I. Tanaka, "First principles phonon calculations in materials science," *Scripta Materialia,* vol. 108, pp. 1-5, 2015.